\documentstyle[12pt]{article}

\author{
~~\\
Re'em SARI\\
Racah Institute for Physics\\
The Hebrew University, Jerusalem 91904, Israel\\
\and
Ramesh NARAYAN\\
Harvard-Smithsonian Center for Astrophysics\\
60 Garden Street, Cambridge MA 02138, U.S.A\\
\and
Tsvi PIRAN\\
Racah Institute for Physics\\
The Hebrew University, Jerusalem 91904, Israel\\
~~\\
}
\title{Cooling Time Scales and Temporal Structure of Gamma-Ray Bursts
}

\input tcilatex

\QQQ{Language}{American English}

\begin{document}

\maketitle

\newpage
\begin{abstract}
A leading mechanism for producing cosmological gamma-ray bursts (GRBs)
is via ultra-relativistic particles in an expanding fireball.  The
kinetic energy of the particles is converted into thermal energy in
two shocks, a forward shock and a reverse shock, when the outward
flowing particles encounter the interstellar medium.  The thermal
energy is then radiated via synchrotron emission and Comptonization.
We estimate the synchrotron cooling time scale of the shocked material
in the forward and reverse shocks for electrons of various Lorentz
factors, focusing in particular on those electrons whose radiation
falls within the energy detection range of the BATSE detectors. We
find that in order to produce the rapid variability observed in most
bursts the energy density of the magnetic field in the shocked
material must be greater than about 1\% of the thermal energy
density. Additionally, the electrons must be nearly in equipartition
with the protons, since otherwise we do not have reasonable radiative
efficiencies of GRBs.  Inverse Compton scattering can increase the
cooling rate of the relevant electrons but the Comptonized emission
itself is never within the BATSE range. These arguments allow us to
pinpoint the conditions within the radiating regions in GRBs and to
determine the important radiation processes.  In addition, they
provide a plausible explanation for several observations.  The model
predicts that the duty cycle of intensity variations in GRB light
curves should be nearly independent of burst duration, and should
scale inversely as the square root of the observed photon energy.
Both correlations are in agreement with observations.  The model also
provides a plausible explanation for the bimodal distribution of burst
durations.  There is no explanation, however, for the presence of a
characteristic break energy in GRB spectra.
\end{abstract}

\centerline{{\it Subject heading: gamma-rays: bursts- 
hydrodynamics-relativity}}

\section{Introduction}

A cosmological gamma-ray burst (GRB) occurs, most likely, in the
deceleration of a shell of ultra-relativistic particles encountering a
surrounding interstellar medium (ISM) (M\'esz\'aros \& Rees 1992).
This process is believed to be essential for the production of a GRB
regardless of the specific nature of the original source of
relativistic particles (see e.g. Piran 1995 for a discussion). In a
recent paper, Sari and Piran (1995, denoted SP hereafter) estimated
the hydrodynamical time scales that arise in the interaction of an
ultra-relativistic shell with the ISM. They showed that the observed
durations of GRBs impose a direct limit on the Lorentz factors of the
relativistic particles, namely $\gamma >100$ for most bursts and
$\gamma$ even larger in a few cases.  SP also worked out the
hydrodynamical conditions in the various fluid zones of the expanding
shell.  They calculated the bulk velocity, thermal energy and particle
density in the shocked material behind the forward and reverse shocks,
as well as the velocities of the two shock fronts.

The radiation we observe in a GRB is produced when the shock-heated
gas loses its thermal energy through various radiation processes.  In
this paper we consider the cooling of the shocked material via
Comptonized synchrotron emission.

In developing a radiation model for GRBs, we can consider two distinct
possibilities, depending on the relative magnitudes of the cooling
time scale of the thermal electrons, $t_{cool}$, and the
hydrodynamical time scale of the expanding shell, $t_{hyd}$.  The case
we focus on in this paper is similar to that assumed by SP, namely
that $t_{hyd}>t_{cool}$.  This assumption finds strong support in the
fact that most bursts have complicated temporal structure with
multiple peaks.  A natural explanation is that the total duration of a
burst is due to $t_{hyd}$, the time needed to convert the bulk of the
kinetic energy of the expanding shell into thermal energy via the two
shocks, while the individual peaks within the profile arise because of
shot-like thermalization events in the shocks.  In this picture, the
individual sub-peaks within a burst represent the cooling curves of
episodically heated electrons.  The width of an individual sub-peak
then represents the cooling time scale $t_{cool}$, and the ``duty
cycle'' of the burst, which we define to be the ratio of the observed
width of an individual peak to the total duration of the burst, is
given by the ratio $D=t_{cool}/t_{hyd}$.

The second case, which we do not consider in this paper, is when
$t_{hyd}<t_{cool}$.  Here, the kinetic energy of the expanding shell
is turned rapidly into thermal energy, but is then released as
radiation on a slower time scale.  Clearly, this regime can only lead
to smooth single hump bursts.  Since most bursts have multiple
sub-peaks, it is necessary to identify each peak with a single hump,
which means that $t_{hyd}$ must be smaller than the width of a
sub-peak.  But such short $t_{hyd}$ requires extremely high Lorentz
factors (SP).  Additionally, a new mechanism is needed to explain the
overall duration of the burst.

In addition to the assumption $t_{hyd}>t_{cool}$, we make the further
assumption that the hydrodynamical conditions do not change
significantly during the cooling of the electrons.  In other words,
we assume that the rapid cooling does not drastically modify the
adiabatic shock structure calculated by SP.  This is a reasonable
approximation if the thermalization in the shocks transfers half or
more of the energy to the protons and only the remainder to the
electrons.  While the electrons cool rapidly, the proton energy
remains locked up in the gas, leaving the hydrodynamical conditions
relatively unaffected.  The true conditions will thus differ from the
idealized adiabatic shocks considered by SP by only factors of order
unity which we ignore.

In this paper, we use two pieces of information from observations to
constrain the parameters of our model of the radiating regions of
GRBs.  First, we note that the vast majority of GRBs have complex time
profiles where individual sub-peaks are significantly narrower than
the overall duration of a burst.  Roughly, the observations give a
mean duty cycle of about 5\%.  As we show, this provides a significant
constraint on the model.  Second, we demand efficient conversion of
thermal energy into radiation at the two shocks since we feel that low
efficiency burst models are implausible.  The observations show that
the net energy emitted by cosmological bursts just within the BATSE
band is $\sim10^{51}~{\rm ergs}$ (Cohen \& Piran 1995, 
Fenimore et. al. 1993).
In the most popular models of bursts, namely merging neutron star
binaries (Narayan, Paczy\'nski \& Piran 1993) and failed supernovae 
(Woosley 1993), the total energy budget is only $\sim10^{53.5}~{\rm
ergs}$, and there are some difficulties in converting even 1\% of the
initial explosion energy into kinetic energy of the expanding shell.
Any further inefficiency in the conversion of shock thermal energy
into radiation would be catastrophic.

We show in this paper that to produce the rapid variability observed
in many bursts the magnetic energy density in the shocked region must
be at least 1\% of the random thermal energy density of the gas.
Additionally, we show that in order to have a reasonable radiative
efficiency, the electrons must be nearly in equipartition with the
protons.  Thus, the observational data on GRBs restrict directly the
conditions in the emitting region of these sources.

We begin the paper with a brief summary in section 2 of the main
results from SP, namely the hydrodynamical time scale of the shell and
the characteristics of the forward and reverse shocks.  We continue
with an estimate of the synchrotron cooling time scale in section 3,
and establish a lower limit for the magnetic field strength in the
shocked material. In section 4 we examine the role of inverse Compton
(IC) emission. We show that this process is irrelevant in the forward
shock since the electrons there are too energetic and the scattering
cross section is reduced considerably according to the Klein-Nishina
formula.  IC can be important in the reverse shock and can increase
the cooling rate there.  However, the IC photons themselves will
generally be outside the energy range of the BATSE detector and
therefore are not relevant for understanding the BATSE data.

In section 5 we discuss the distribution of electron energy and the
effect this has on the ``efficiency'' of a burst, namely the fraction
of the initial kinetic energy in the shell which finally appears as
radiation in the BATSE energy window.  We show that the forward shock
often produces nearly all its radiation at energies above the BATSE
range, especially for short duration bursts.  In contrast, the reverse
shock is always visible to BATSE.  However, the efficiency of the
reverse shock is usually somewhat low, whereas the forward shock, when
it radiates within the BATSE window, always has a high efficiency.  We
show in section 6 that these characteristics of the two shocks provide
a plausible explanation for the bimodal distribution of GRB durations
observed by BATSE (Kouveliotou et. al. 1993, Lamb, Graziani and Smith,
1993).  We suggest that short bursts originate from the reverse shocks
of fireballs which expand with high Lorentz factor $\gamma$, while
long bursts originate from the forward shock of low $\gamma$ events.
This simple model is also in agreement with the relative luminosities
of short and long bursts as estimated by Mao, Narayan \& Piran (1994).

\section{Shock Conditions and Hydrodynamic Time Scales}

We begin with a brief summary of the hydrodynamic conditions and
energy conversion time scales of a relativistically expanding shell of
particles, treated as a fluid (see SP and Piran 1995 for further
details). The interaction between the outward moving shell and the ISM
takes place in the form of two shocks: a forward shock that propagates
into the ISM and a reverse shock that propagates into the relativistic
shell. This results in four distinct regions: the ISM at rest (denoted
by the subscript 1 when we consider properties in this region), the
shocked ISM material which has passed through the forward shock
(subscript 2 or f), the shocked shell material which has passed
through the reverse shock (3 or r), and the unshocked material in the
shell (4). See Figure 1.

The hydrodynamic conditions are determined by four parameters: the
Lorentz factor $\gamma$ associated with the relativistic radial
expansion of the shell of particles, the width of the shell in the
observer frame $\Delta $, the density of the external ISM $n_1$, and
the Sedov scale $l=(E/n_1m_pc^2)$, where $E$ represents the total
energy of the shell before interaction with the ISM.  Typical values
of these parameters are: $\gamma\sim10^2-10^4$, $\Delta<10^{13}$ cm
(from the fact that bursts are almost always less than a few hundred
seconds long, see SP), $n_1=1~{\rm particle\;cm^{-3}}$, and
$l=10^{18}$cm (corresponding to a burst with energy
$E=1.5\times10^{51}$ergs).

The radial expansion speeds of the four zones are described by the
following Lorentz factors.  The unshocked ISM is of course at rest and
has $\gamma_1=1$, while the unshocked shell material moves at the
original coasting velocity of the particles, $\gamma_4=\gamma$.  The
velocity of the two intermediate zones satisfies
$\gamma_1<\gamma_2=\gamma_3<\gamma_4$.  The relative Lorentz factor
across the forward shock is obviously equal to $\gamma_2$.  The
relative Lorentz factor across the reverse shock, which we write as
$\bar\gamma_3$, is given in the ultra-relativistic limit by
\begin{equation}
\bar\gamma_3=\gamma/\gamma_3,
\end{equation}
where we have ignored, as in the rest of the paper, factors of order
unity.

SP showed that the structure of the shocks depends on the value of
$\bar \gamma _3$, which in turn depends on the parameter $\xi$:
\begin{equation} 
\label{xidef}
\xi \equiv{{\left( {l\over\Delta }\right)}^{1/2}\gamma ^{-4/3}}.  
\end{equation} 
If $\xi \gg 1$ the reverse shock is non-relativistic or Newtonian, and
$\bar\gamma_3\sim1$, while if $\xi \ll 1$ the reverse shock is
relativistic and $\bar\gamma_3\gg1$.  If we include the possibility of
shell spreading (see SP for details), then $\Delta $ changes with time
in such a manner that at each moment the current $\Delta (t)\sim {\rm
max}(\Delta (0),R/\gamma ^2)$.  This means that a shell which begins
with a value of $\xi>1$ adjusts itself so as to satisfy $\xi=1$ and we
have a mildly relativistic or Newtonian reverse shock, whereas a shell
with $\xi<1$ does not have time to spread significantly and maintains
a relativistic reverse shock.  In the rest of the paper, we
concentrate on the relativistic case and express all results in terms
of the parameter $\xi$ (rather than $\Delta $).  By setting $\xi<1$ in
the expressions we obtain results corresponding to a relativistic
reverse shock, and by choosing $\xi=1$ in the same expressions we
obtain the spreading Newtonian limit. We shall not discuss the case of
extreme Newtonian reverse shock ($\xi\gg1$), since spreading will
always bring these shells to the mildly relativistic limit
($\xi\sim1$)which we call ``Newtonian'' in the rest of the paper.
Therefore, the same formulae are valid in both the relativistic and
Newtonian limits.

According to the arguments presented in the Introduction, the observed
duration $t_{dur}$ of a burst is given by the hydrodynamic time
$t_{hyd}$ of the shell, which has been calculated by SP to be
\begin{equation}
\label{tnr}t_{dur}=\left({l\over c}\right)\gamma ^{-8/3}\xi ^{-2}=
(150~{\rm s})\left({\gamma\over100}\right)^{-8/3}\xi ^{-2}l_{18},
\end{equation}
where $l_{18}=l/10^{18}$ cm.  This time scale is equal to $\Delta/c$
in both the relativistic and Newtonian cases, except that in the
former case the width is independent of the radius of the shell
whereas in the latter the width self-consistently expands so as to
maintain $\xi=1$.

The bulk of the kinetic energy of the shell is converted to thermal
energy via the two shocks at around the time the shell has expanded
to the radius $R_\Delta \equiv l\gamma ^{-2/3}\xi ^{-1/2}$. At this
radius, the conditions at the forward shock are as follows,
\begin{eqnarray}
\label{hydroforward}
\gamma _2 & = & \gamma \xi ^{3/4},  \nonumber \\
n_2 & = & 4\gamma _2n_{1},  \\
e_2 & = & 4\gamma _2^2n_{1}m_pc^2, \nonumber
\end{eqnarray}
while at the reverse shock we have
\begin{eqnarray}
\label{hydroreverse}
\bar \gamma _3 & = & \xi^{-3/4},  \nonumber \\
\gamma_3       & = & \gamma\xi^{3/4}, \\
n_3            & = & 4\xi ^{9/4}\gamma ^2n_{1}, \nonumber \\
e_3            & = & e_2. \nonumber
\end{eqnarray}
We see from equation (\ref{tnr}) that the observed duration of a burst
depends on $\gamma$ and $\xi$ only through the combination
$\gamma_2=\gamma\xi^{3/4}$.  Therefore many observed quantities which
depend on $\gamma_2$ can be expressed in terms of $t_{dur}$ via the
relation (see equations \ref{tnr} and \ref{hydroforward})
\begin{equation} 
\label{g2}\gamma_2=280\left( \frac
{t_{dur}} {10~{\rm s}} \right)^{-3/8} l_{18}^{3/8}.
\end{equation}

The efficiency of the cooling processes and the nature of the emitted
radiation depend on the conditions in the shocked regions 2 and 3.
Both regions have the same energy density $e$. The particle densities
$n_2$ and $n_3$ are, however, different and hence the effective
``temperatures,'' i.e. the mean Lorentz factors of the random motions
of the shocked protons and electrons, are different.  The parameters
that determine the radiative cooling are the magnetic field strength,
$B$, and the distribution of electron Lorentz factor $\gamma_e$.  Both
of these are difficult to estimate from first principles.  Our
approach is to use two parameters, $\epsilon_B$ and $\epsilon_e$
defined below, to incorporate our uncertainties.  We then explore the
space of these parameters and try to constrain the parameter values by
requiring the model predictions to resemble BATSE observations of
GRBs.

The dimensionless parameter $\epsilon_B$ measures the ratio of the
magnetic field energy density to the total thermal energy $e$:
\begin{equation}
\label{EpsilonB} \epsilon_B\equiv {U_B\over e} = {B^2\over8\pi e}, 
\end{equation}
so that
\begin{equation}
B=4 \sqrt{2\pi} c \epsilon_B^{1/2} \gamma_2 m_p^{1/2} n_{1}^{1/2}
=(110~{\rm G})\epsilon_B^{1/2}\left({t_{dur}\over 10~{\rm s}}\right)
^{-3/8}l_{18}^{3/8}n_{1}^{1/2}.
\end{equation}
If the magnetic field in region 2 behind the forward shock is obtained
purely by shock compression of the ISM field, the field would be very
weak, with $\epsilon_B \ll 1$.  Such low fields are incompatible with
observations of GRBs as we show in later sections.  We therefore
consider the possibility that there may be some kind of a turbulent
instability which may bring the magnetic field to approximate
equipartition.  In the case of the reverse shock, magnetic fields of
considerable strength might be present in the pre-shock shell material
if the original exploding fireball was magnetic. The exact nature of
magnetic field evolution during fireball expansion depends on several
assumptions. Thompson (1994) found that the magnetic field will remain
in equipartition if it started off originally in equipartition.
M\'esz\'aros, Laguna \& Rees (1993) on the other hand estimated that
if the magnetic field was Initially in equipartition then it would be
below equipartition by a factor of $10^{-5}$ by the time the shell
expands to $R_\Delta$.  As in the forward shock, an instability could
boost the field back to equipartition.  Thus, while both shocks may
have $\epsilon_B\ll1$ with pure flux freezing, both could achieve
$\epsilon_B\rightarrow1$ in the presence of instabilities.  In
principle, $\epsilon_B$ could be different for the two shocks, but we
limit ourselves to the same $\epsilon_B$ in both shocks.

Our second parameter $\epsilon_e$ measures the fraction of the total
thermal energy $e$ which goes into random motions of the electrons:
\begin{equation}
\epsilon _e\equiv {U_e\over e}. 
\end{equation}
We assume that the thermal energy immediately behind the shock is
divided up in the ratio $(1-\epsilon_e):\epsilon_e$ between the
protons and the electrons.  We assume that the division happens on a
time-scale which is shorter than the electron cooling time, and
therefore much shorter than the hydrodynamic time.  We further assume
that once the initial allocation of the energy has been accomplished,
no further energy flows to the electrons from the protons.  (It is
easily seen that Coulomb coupling is completely negligible at the
densities of interest.)

Since the electrons receive their random motions through
shock-heating, we make the standard assumption that they develop a
power law distribution of Lorentz factors,
\begin{equation}
N(\gamma_e )\sim \gamma_e ^{\bar \beta }\ \ \ {\rm for}\ \gamma_e
>\gamma _{e,min}\;.
\end{equation}
We require $\bar \beta <-2$ so that the energy does not diverge at
large $\gamma_e$.  Since our shocks are relativistic we assume that
all the electrons participate in the power-law, not just a small
fraction in the tail of the distribution as in the Newtonian case.
The minimum Lorentz factor, $\gamma _{e,min}$, of the distribution is
related to $\epsilon _e$ by 
\begin{equation} \label{epsilone}
\epsilon_ee=\gamma_{e,min}nm_ec^2\frac{\bar \beta +1 }{\bar \beta
+2},\qquad e\sim\gamma_{sh}nm_pc^2, 
\end{equation} 
where $\gamma_{sh}$ is the relative Lorentz factor across the shock
front.  Thus, 
\begin{equation}
\gamma_{e,min}=1840{(\bar\beta+2)\over(\bar\beta+1)}\epsilon_e\gamma_{sh}.
\end{equation} 
The average $\gamma_e$ of the electrons is given by
\begin{equation}
\langle\gamma_e\rangle={(\bar \beta +1)\over(\bar \beta
+2)}\gamma _{e,min}\;.
\end{equation}
Note that $\gamma_{sh}=\gamma_2$ for the forward shock and
$\gamma_{sh} =\bar\gamma_3$ for the reverse shock.

The energy index $\bar\beta$ can be fixed by requiring that the model
should be able to explain the high energy spectra of GRBs.  If we
assume that most of the radiation observed in the BATSE window is due
to synchrotron cooling (as we confirm later in the paper), then it is
straightforward to relate $\bar\beta$ to the power-law index of the
observed spectra of GRBs.  Band et al. (1993) measured the mean
spectral index of GRBs at high photon energies (above the break) to be
$\beta\approx-2.25$, which corresponds to $\bar\beta\approx-2.5$.  We
assume this value of $\bar\beta$ in what follows:
\begin{equation}
\bar\beta=-2.5,\qquad {(\bar\beta+1)\over(\bar\beta+2)}=3\;.
\end{equation}
The electrons in zone 2 behind the forward shock then satisfy
\begin{equation}
\label{gammaeminf}
\gamma_{e,min}={1\over3}\langle\gamma_e\rangle=
610\epsilon_e\gamma_2=1.7\times10^5\epsilon_e
\left({t_{dur}\over10~{\rm s}}\right)^{-3/8}l_{18}^{3/8},
\end{equation}
while the electrons in zone 3 behind the reverse shock have
\begin{equation}
\label{gammaeminr}
\gamma_{e,min}={1\over3}\langle\gamma_e\rangle=610\epsilon_e\xi^{-3/4}.
\end{equation}
We shall see later that the precise details of the distribution of
$\gamma_e$ are not very important for most of the calculations
presented in this paper; only the value of $\gamma_{e,min}$ is
relevant.

\section{Synchrotron Cooling}

The typical energy of synchrotron photons as well as the synchrotron
cooling time depend on the Lorentz factor $\gamma_e$ of the
relativistic electrons under consideration and on the strength of the
magnetic field. The characteristic photon energy in the fluid frame is
given by
\begin{equation}
\label{hnufluid}
(h\nu_{syn})_{fluid}=\frac{\hbar q_eB}{m_ec}\gamma _e^2 . 
\end{equation}
Since the emitting material in both shocks moves with a Lorentz
factor $\gamma_2$ the photons are blue shifted in the observer frame:
\begin{equation}
\label{ge}(h\nu_{syn})_{obs}=\frac{\hbar q_eB}{m_ec}\gamma _e^2\gamma_2 
=\left(3.5\times10^{-7}~{\rm KeV}\right)\gamma_e^2\epsilon_b^{1/2}
\left({t_{dur}\over10~{\rm s}}\right)^{-3/4}l_{18}^{3/4}n_1^{1/2}.
\end{equation}

The power emitted by a single electron due to synchrotron radiation
(see e.g.  Rybicki and Lightman, 1979) is:
\begin{equation}
P = \frac 4 3 \sigma_T c U_B \gamma_e^2 \ , 
\end{equation}
where $\sigma_T$ is the Thomson cross section.  The cooling time of
the electron in the fluid frame is then $\gamma_e m_e c^2/P$. The
observed cooling time $\tau_{syn}$ is shorter by a factor of
$\gamma_2$, giving
\begin{equation}
\label{cooling}\tau_{syn}=\frac{3\gamma _em_ec^2}{4\sigma _Tc
U_B\gamma_e^2\gamma_2}=(230~{\rm s}){1\over\gamma_e}\epsilon_B^{-1}
\left({t_{dur}\over10~{\rm s}}\right)^{9/8}l_{18}^{-9/8}n_1^{-1}.
\end{equation}

The above results depend on the choice of $\gamma_e$.  One possibility
(following M\'esz\'aros, Laguna \& Rees, 1993) is to consider the
Lorentz factor of a ``typical electron'' in the emitting region,
i.e. to set $\gamma_e=\langle\gamma_e\rangle$, and to use this for
estimating the cooling time scale. However, it is not clear that the
photons emitted by such electrons will actually be within BATSE's
range.  Since BATSE detects photons with energies $\sim 100$ KeV, we
concentrate on electrons with Lorentz factor $\hat \gamma _e$,
where $\hat\gamma_e$ is that Lorentz factor at which an electron emits
synchrotron photons with energies around $\sim 100$ KeV.  Of course,
other electrons are also present in the medium.  But the electrons
with lower energies emit softer photons while the higher energy
electrons emit harder photons which fall outside the BATSE range (but
may be detected by other experiments which are more sensitive to these
photons).  In view of our interest in understanding the BATSE
observations we concentrate on the behavior of the electrons with
$\gamma _e\approx \hat \gamma _e$.  We do retain a scaling factor in
terms of the observed photon energy $h\nu_{obs}$ in the results so
that the cooling times of other electrons are also implicit in our
relations.

Using, Eq. \ref{ge} we calculate $\hat \gamma _e$ to be given by
\begin{equation}
\label{gamma_hat}
\hat \gamma _e=\left( \frac{m_ech\nu_{obs}}
{\hbar q_e\gamma _2B}\right) ^{1/2} = 1.7\times10^4\epsilon_B^{-1/4}
\left({h\nu_{obs}\over100~{\rm KeV}}\right)^{1/2}
\left({t_{dur}\over10~{\rm s}}\right)^{3/8}l_{18}^{-3/8}n_1^{-1/4}.
\end{equation}
We first need to check that electrons with $\gamma_e=\hat\gamma_e$ are
available in the shocked material.  This means that we require
$\gamma_{e,min}<\hat \gamma_e$, which corresponds to the condition
\begin{equation}
\label{maxeer} \epsilon _{e~|r}<28\epsilon _b^{-1/4}
\left({h\nu_{obs}\over100~{\rm KeV}}\right)^{1/2}
\left({t_{dur}\over10~{\rm s}}\right)^{3/8}\xi^{3/4}l_{18}^{-3/8}n_1^{-1/4}
\end{equation}
in the reverse shock, and the condition
\begin{equation}
\label{maxeef} \epsilon _{e~|_f}<0.1\epsilon _B^{-1/4}
\left({h\nu_{obs}\over100~{\rm KeV}}\right)^{1/2}
\left({t_{dur}\over10~{\rm s}}\right)^{3/4}l_{18}^{-3/4}n_1^{-1/4}
\end{equation}
in the forward shock.  Since by definition $\epsilon_e\leq1$, we see
that the reverse shock always has electrons with the right Lorentz
factors to produce synchrotron photons within the BATSE range.
However, the situation is more doubtful in the case of the forward
shock.  If the heating of the electrons is efficient, i.e. if
$\epsilon_{e~|_f}\sim1$, and if the burst has a short duration, then
most of the electrons may be too energetic to produce BATSE-visible
photons.
Of course, as an electron cools, it radiates at progressively softer
energies.  Therefore, even if $\gamma_{\rm min}$ is initially too
large for the synchrotron radiation to be visible to BATSE, the same
electrons would at a later time have $\gamma_e\sim\hat\gamma_e$ and become
visible to BATSE.  However, the energy remaining in the electrons at
the later time will also be lower (by a factor $\hat\gamma/\gamma_{\rm
min}$), which means that the burst will be inefficient.  For
simplicity, we ignore this radiation.

Substituting the value of $\hat \gamma_e$ from equation
\ref{gamma_hat} into the cooling rate equation \ref{cooling} we
obtain the cooling time scale as a function of the observed photon
energy to be
\begin{equation}
\label{tausyn2}\tau _{syn}=(1.4\times 10^{-2}~{\rm s}) \epsilon_B^{-3/4}
\left({h\nu_{obs}\over100~{\rm KeV}}\right)^{-1/2}
\left({t_{dur}\over10~{\rm s}}\right)^{3/4}l_{18}^{-3/4}n_1^{-3/4}.
\end{equation}
Equation \ref{tausyn2} is valid for both the forward and reverse
shock, and is moreover independent of whether the reverse shock is
relativistic or Newtonian.  It is therefore quite a robust result.

The cooling time calculated above sets a lower limit to the
variability time scale of a GRB since the burst cannot possibly
contain spikes that are shorter than its cooling time.  Observations
of GRBs typically show asymmetric spikes in the intensity variation,
where a peak generally has a fast rise and a slower exponential
decline (FRED).  A plausible explanation of this observation is that
the shock heating of the electrons happens rapidly (though
episodically), and that the rise time of a spike is related to the
heating time.  The decay time is then set by the cooling, so that the
widths of spikes directly measure the cooling time.

We see from Eq. \ref{tausyn2} that $\tau _{syn}$ is proportional to
$\left( h\nu\right) ^{-1/2}$.  This suggests that there should be an
inverse correlation between the width of a spike and the energy band
in which the observation is made.  An inverse correlation has indeed
been observed (Fenimore 1995).  In fact, even the predicted dependence
as the inverse square root of the photon energy is quite close to the
observed index of $-0.4$.

Using Eqs. \ref{tausyn2} and \ref{tnr} we can estimate the duty-cycle
$D$ of the variability as the ratio of the cooling time scale to the
total duration:
\begin{equation}
\label{dutycyclesyn} 
D\equiv{\frac{\tau _{syn}}{t_{dur}}}=1.4\times10^{-3}\epsilon_B^{-3/4}
\left({h\nu_{obs}\over100~{\rm KeV}}\right)^{-1/2}
\left({t_{dur}\over10~{\rm s}}\right)^{-1/4}l_{18}^{-3/4}n_1^{-3/4}.
\end{equation} 
We see that the duty cycle depends only weakly on the burst duration.
Observationally, it appears to be true that the narrowest features in
a burst are a constant fraction of the total duration, independent of
the actual duration of a burst (e.g. Bhat 1994), but this prediction
of the model needs to be checked in more detail against
observations.

Most GRBs are seen to be highly variable with duty-cycles
significantly smaller than unity, typically on the order of
$5\%$ or less.  Using this as a constraint, we obtain
a lower limit to the value of $\epsilon _B$:
\begin{equation}
\label{lowerlimitsyn}\epsilon _B\ge 8.4 \times 10^{-3}
\left({D\over0.05}\right)^{-4/3}
\left({h\nu_{obs}\over100~{\rm KeV}}\right)^{-2/3}
\left({t_{dur}\over10~{\rm s}}\right)^{-1/3}l_{18}^{-1}n_1^{-1}.
\end{equation}
This suggests that the magnetic field energy density cannot be far
less than the equipartition value in the radiation emitting regions
in GRBs. Clearly, the limit is valid only if synchrotron radiation is
the main cooling process. In particular we have ignored so far the
inverse Compton process which could increase the emission and thereby
allow a lower value of $\epsilon_B$.  We turn now to the implications
of inverse Compton scattering.

\section{Inverse Compton Scattering}

Inverse Compton (IC) scattering may modify our analysis in three ways. 

First, a significant fraction of the synchrotron emission may be
scattered so that the observed synchrotron flux is lower than the
calculated value.  In fact, this is never the case because we can show
that the emitting regions are always extremely optically thin.

Second, IC might scatter low energy synchrotron photons into the
BATSE energy band, so that some of the observed radiation may be due
to IC rather than synchrotron emission. We show that this is again
not the case. IC photons from the forward shock are always much
harder than the BATSE range.  Although IC photons from the reverse
shock can fall within the BATSE range, they do so only for bursts
with large duty cycles, i.e. for smooth one-hump bursts which form
only a minority of the observed bursts.

Finally, even if IC does not influence any of the observed photons it
may speed up the cooling of the emitting regions and therefore shorten
the cooling time which we estimated earlier under the assumption of
pure synchrotron emission.  This effect is unimportant for the forward
shock but could be important for the reverse shock under certain
conditions.

Let us begin with the optical depth.  The radial thicknesses of
zones 2 and 3 are each of order $\gamma_2\Delta$ in the comoving frame.
Therefore the Thomson optical depth across the full radial extent of
both zones is given by 
\begin{equation} 
\tau =(n_3\gamma _2\Delta
+n_2\gamma _2\Delta )\sigma _T = 4\times 10^{-6}
\left({t_{dur}\over10~{\rm s}}\right)^{-1/8}\xi^{-3/4}n_1,
\end{equation} 
where we have used the inequality $n_2 \ll n_3$.  In fact this is an
overestimate of $\tau$ since some of the electrons are very energetic
and have a reduced scattering cross-section by the Klein-Nishina
effect.  Even without allowing for this effect we see that the optical
depth is extremely small for reasonable parameters.

Considering the second point, synchrotron photons emitted by
electrons of Lorentz factor $\gamma_{e,1}$ and inverse Compton
scattered by electrons of $\gamma_{e,2}$ 
have an observed energy of
\begin{equation}
\label{hnuic}(h\nu _{IC})_{obs}=\frac{\hbar q_eB}{m_ec}\gamma _{e,1}^2
\gamma_{e,2}^2\gamma_2= 350~{\rm MeV}
\left({\gamma_{e,1}\over10^3}\right)^2
\left({\gamma_{e,2}\over10^3}\right)^2 \epsilon_B^{1/2}
\left({t_{dur}\over10~{\rm s}}\right)^{-3/4}l_{18}^{3/4}n_1^{1/2}.
\end{equation}
For a photon to be detectable in BATSE, its observed energy has to be
in the range 25 KeV to about 1 MeV.  This requires $\gamma_{e,1},
\gamma_{e,2}<200$.  Now, if we consider an efficient burst with
$\epsilon_e\sim1$, then the typical value of $\gamma_{e,min}$ is
$\gg1000$ for the forward shock, and is $>1000$ even for the reverse
shock if it is relativistic.  Only for a Newtonian reverse shock
(where $\xi\sim1$) with somewhat low efficiency ($\epsilon_e<0.3$) is
$\gamma_{e,min}< 1000$.  Stating this differently, the condition
$\gamma _{e,min}<200$ yields, using Eq. \ref{epsilone}, an extremely
small $\epsilon _e$ for the forward shock:
\begin{equation}
\epsilon _e<7.6\times 10^{-4}\epsilon_B^{-1/8}
\left({h\nu_{obs}\over100~{\rm KeV}}\right)^{1/4}
\left({t_{dur}\over10~{\rm s}}\right)^{9/16}l_{18}^{-9/16}n_1^{-1/8}.
\end{equation}
while for the reverse shock we get
\begin{equation}
\label{epsilonmaxic}\epsilon _e<0.21\epsilon_B^{-1/8}
\left({h\nu_{obs}\over100~{\rm KeV}}\right)^{1/4}
\left({t_{dur}\over10~{\rm s}}\right)^{3/16}\xi^{3/4}l_{18}^{-3/16}n_1^{-1/8}.
\end{equation}
IC photons from the forward shock are thus completely out of the range
of BATSE unless $\epsilon_e$ is extremely small, a possibility which
we eliminate on the grounds of inefficiency (see the discussion in the
Introduction).  With the reverse shock, the best chance is with $\xi$
equal to its largest value, namely $\xi=1$, and considering a long
burst duration.  We show below that even this case results in a long
cooling time and a large duty cycle and therefore is ruled out by the
observed peaky nature of bursts.

We next turn to the question of whether IC losses can significantly
alter the cooling time scale of the electrons with Lorentz factor
$\hat \gamma _e$ which emit the synchrotron radiation observed by
BATSE.  Because of the Klein-Nishina effect, the cross section for
Compton scattering decreases quite rapidly when the energy of the
synchrotron photons becomes larger than $m_e c^2$ in the electron rest
frame. IC is therefore most important when the synchrotron photons are
below this Klein-Nishina cut-off.  We simplify matters by assuming
that IC switches off completely once the Klein-Nishina regime is
entered.  This is of course a rather severe simplification, but we
consider it justified for the purposes of the present argument.  The
lowest energy synchrotron photons available are those photons emitted
by the lowest energy electrons, with $\gamma_{e1}=\gamma_{e,min}$
(Eq. \ref {epsilone}). The IC process can be neglected if even these
low energy photons are in the Klein-Nishina range for an electron of
Lorentz factor $\hat \gamma_e$, i.e. if
\begin{equation}
(h\nu_{min})_{fluid} \hat \gamma _e \ge m_ec^2 . 
\end{equation}
Using the formula for $(h\nu)_{fluid}$ given in equation
(\ref{hnufluid}) and setting $\gamma_e=\gamma_{e,min}$
(Eq. \ref{gammaeminf} or \ref{gammaeminr}) and substituting the values
of $e$ and $n$ (Eqs. \ref {hydroforward} or \ref{hydroreverse}), we
find that IC can be neglected if
\begin{equation}
\epsilon _e > 2.9 \times 10^{-2} \epsilon_B^{-1/8} 
\left( \frac{h\nu_{\rm obs} }{100\text{KeV}}
\right)^{-1/4}\left( \frac{t_{dur}}{10~\text{s}}\right)^{3/8}
l_{18}^{-3/8}n_1^{-1/8} ,
\end{equation}
for the forward shock, and if
\begin{equation}
\epsilon _e>8.0 \epsilon_B^{-1/8} 
\left( \frac{h\nu_{\rm obs} }{100\text{KeV}}\right)^{-1/4}
\xi^{3/4} n_1^{-1/8},
\end{equation}
for the reverse shock.

Since on the grounds of efficieny we have argued that $\epsilon_e$
must be close to unity, we see that the condition in the case of the
forward shock is almost always satisfied and therefore IC cooling is
never important there.  The reverse shock, on the other hand, never
satisfies the condition and so IC cooling might be important.

Since IC is not important for the forward shock, the relations
(\ref{dutycyclesyn}) and (\ref {lowerlimitsyn}) derived in section 3
are valid without any change.  However, for the reverse shock, we need
to calculate the true cooling time including the additional cooling
due to IC.  To do this we use the following model.  The energy density
available for conversion to radiation is the electron energy density,
$U_e$.  Eventually all this energy is radiated either by synchrotron
or by IC.  We denote the former by $U_{syn}$ and the latter by
$U_{IC}$:
\begin{equation}
\label{ic1} U_e=U_{syn}+U_{IC} 
\end{equation}

If the shock has already propagated for a period of time longer than
the cooling time we expect a steady state radiation density profile to
be present in the vicinity of the shock-heated electrons.  In this
steady state, we expect the energy flux given by the shock to the
electrons to be equal to the radiation flux.  Further, since the shock
front moves with a velocity equal to a fraction of the speed of light,
the total radiation density must be comparable to $U_e$, with the
synchrotron and IC densities given respectively by $U_{syn}$ and
$U_{IC}$.

Because of the Klein-Nishina effect, we need consider only one
scattering of each photon, and further scatterings are highly
suppressed.  In this limit, the Compton $y$ parameter is directly
equal to the ratio of the rate of IC energy loss to the rate of
synchrotron energy loss.  At the same time, $y$ is also equal to the
ratio of the energy density in soft radiation (in our case
synchrotron) to magnetic field energy density (Rybicki \& Lightman
1979).  We therefore find 
\begin{equation}
\label{ic2}
y=\frac{U_{IC}}{U_{syn}}=\frac{U_{syn}}{U_B} . 
\end{equation}
Recalling that $U_e=\epsilon_ee$ and $U_B=\epsilon_Be$, we can solve
equations \ref{ic1} and \ref{ic2} to obtain
\begin{equation}
y\equiv \frac{U_{IC}}{U_{syn}}={\epsilon _e\over\epsilon _B}\text{ if }
{\epsilon_e\over\epsilon _B}\ll \text{1} \;,
\end{equation}
\begin{equation}
y\equiv \frac{U_{IC}}{U_{syn}}=\sqrt{\epsilon _e\over\epsilon _B}\text{ if }
{\epsilon _e\over\epsilon _B}\gg 1 \;.
\end{equation}

If $\epsilon _e<\epsilon _B$ we see that $y<1$ and IC cooling is not
important.  In this case, we can use equations (\ref{dutycyclesyn})
and (\ref {lowerlimitsyn}) even for the reverse shock.  However, if
$\epsilon_e>\epsilon _B$ then $y\sim\sqrt{\epsilon _e/\epsilon _B}>1$
and most of the emitted radiation comes out as IC photons. The cooling
rate of the relevant electrons then increases by the factor of $y$ and
the lower limit to the magnetic energy density set by the observed
duty-cycle decreases.  Equations (\ref{dutycyclesyn}) and
(\ref{lowerlimitsyn}) are thus replaced by
\begin{equation}
\label{dutycycleic} 
D={\frac{\tau _{syn}}{t_{dur}}}=1.4\times10^{-3}
\epsilon_e^{-1/2} \epsilon_B^{-1/4}
\left({h\nu_{obs}\over100~{\rm KeV}}\right)^{-1/2}
\left({t_{dur}\over10~{\rm s}}\right)^{-1/4}l_{18}^{-3/4}n_1^{-3/4}.
\end{equation} 
\begin{equation}
\label{lowerlimitic} \epsilon_B \ge 6 \times 10^{-7} \epsilon_e^{-2}
\left({D\over0.05}\right)^{-4}
\left({h\nu_{obs}\over100~{\rm KeV}}\right)^{-2}
\left({t_{dur}\over10~{\rm s}}\right)^{-1}l_{18}^{-3}n_1^{-3}.
\end{equation}
We emphasize that these estimates are relevant only for the
reverse shock and only if $\epsilon_e>\epsilon_B$.  For all other
cases, we use equations (\ref{dutycyclesyn}) and (\ref
{lowerlimitsyn}).

Although the duty cycle argument now appears to allow a much lower
magnetic energy density, $\epsilon_B$ as low as $10^{-6}$, we obtain
another more stringent limit by demanding that the burst be efficient.
Since the IC photons are too hard to be observed by BATSE the observed
efficiency decreases by a factor $1/y=\sqrt{\epsilon _B/\epsilon _e}$.
Thus, even if we take the maximum efficiency for the shock
acceleration of electrons, $\epsilon_e=0.5$, still we find that a
value of $\epsilon_B\sim10^{-6}$ results in a burst with an efficiency
$<10^{-3}$ just from this effect.  In fact, the efficiency is a little
lower for reasons described in the next section.  As we explained in
the Introduction, GRB scenarios become quite implausible at such low
efficiencies as they would imply unreasonably large energies in the
original explosion.

We are now ready to show, as we have stated at the beginning of this
section, that we cannot have a highly variable burst of IC photons from the
reverse shock. The Lorentz factor of electrons emitting IC photons with
energy $h \nu$ is given by equation (\ref{hnuic}) as: 
\begin{equation}
\label{gamma_eic} \gamma _e=\left( \frac{m_e c h\nu}{\hbar q_e\gamma _2B}%
\right) ^{1/4} . 
\end{equation}
The corresponding duty-cycle for these electrons is: 
\begin{equation}
\label{dutycycleicobs}\frac{\tau _{syn}}{t_{dur}}=0.18\left( \frac{t_{dur}}{%
10\text{sec}}\right) ^{-1/16}\epsilon _B^{-3/8}\epsilon _e^{-1/2}\left( 
\frac{h\nu }{100\text{KeV}}\right) ^{-1/4}l_{18}^{-15/16}n_1^{-7/8} . 
\end{equation}
As we have already seen, electrons with Lorentz factor given by
Eq. \ref{gamma_eic} are impossible in the forward shock but are
possible in a Newtonian reverse shock provided $\epsilon_e<0.2$.
However, now we see from the above equation that even if the electrons
are present, their radiation will come out with a somewhat large
duty-cycle $\sim1$.  This is ruled out by the observations.

We note that M\'esz\'aros, Laguna \& Rees (1993) choose as the
canonical values for their models, $\epsilon _B=10^{-5}$ and $\epsilon
_e \approx 1$.  The duty-cycle for this choice of parameters is $\sim
10$, which means that the hydrodynamic time is an order of magnitude
shorter than the cooling time.  This runs into problems on two counts.
First, the model can only produce smooth single hump bursts and cannot
explain the variability of observed bursts without invoking
sub-structure within the shell.  Second, the hydrodynamical conditions
in the expanding shell, in particular $\gamma_2$, change on the
hydrodynamic time scale.  This means that only during the first
$t_{hyd}$ of cooling will the radiation be fully beamed with the
Lorentz factor $\gamma_2$.  The later cooling will be less beamed and
will therefore be smeared out significantly in observer time.  What
this means is that for the observer, the burst will effectively last
only for a duration $t_{hyd}$; the remaining 90\% of the energy will
come much later and will be weaker and softer and will not be counted
as part of the burst.  As a result, the observed efficiency of the
burst will be reduced by a factor of order 10.

\section{Efficiency}

We now consider the efficiencies with which the two shocks convert
their thermal energy into observable radiation.  Two important
factors influence the efficiency and lower it from the ideal value of
unity. First, only the electrons cool, which means that only a
fraction $\epsilon_e$ of the total thermal energy is available to be
radiated.  Second, a significant fraction of the radiated energy may
lie outside the BATSE energy range because (1) the synchrotron
emission is usually spread over a large energy range, and (2) much of
the cooling may happen via IC scattering whose radiation is quite
generally at energies above the BATSE range.

We have assumed a power-law distribution of electron energy with a
power-law index $\bar\beta=-2.5$.  This index has been chosen such
that the time-integrated synchrotron emission corresponds to a
power-law radiative spectrum with index $\beta=-2.25$, in agreement
with the observations (Band et al. 1993).  The fraction of the
synchrotron power which is radiated in the BATSE band (at around 100
KeV) is then given by
\begin{equation}
\label{eficeincydis}
\epsilon_{syn}=
\left( \frac{h\nu _{\min }}{100~\text{KeV}}\right) ^{-\beta-2}
= \left( \frac{h\nu _{\min }}{100~\text{KeV}}\right) ^{0.25},
\end{equation}
where $h\nu_{\min }$ is the observed photon energy corresponding to 
the synchrotron radiation from electrons with $\gamma_e=\gamma_{e,min}$.  

For the forward shock, we can calculate $h\nu_{min}$ using equation
\ref{ge} and setting $\gamma_e=\gamma_{e,min}$ from equation 
\ref{gammaeminf}:
\begin{equation}
{h\nu_{min}\over100~{\rm KeV}}=
100\epsilon_e^2\epsilon_B^{1/2}
\left({t_{dur}\over10~{\rm s}}\right)^{-3/2}l_{18}^{3/2}n_1^{1/2}.
\end{equation}
Since IC scattering is unimportant for the forward shock, the net
efficiency is just the product of $\epsilon_e$, the fraction of the
energy which goes into the electrons, and the synchrotron efficiency
factor $\epsilon_{syn}$ given above.  Thus we find the total efficiency
of the forward shock to be
\begin{equation}
\label{effiforward}
\epsilon_{tot,f}=
\epsilon_e\epsilon_{syn}=3.2\epsilon _e^{3/2}\epsilon _B^{1/8}
\left({t_{dur}\over10~{\rm s}}\right)^{-3/8}l_{18}^{3/8}n_1^{1/8}.
\end{equation}
The efficiency is of order unity whenever the forward shock is
visible in the BATSE range (which requires long durations as 
discussed in section 3.)

For the reverse shock we have
\begin{equation}
{h\nu_{min}\over100~{\rm KeV}}=
1.3\times10^{-3}\epsilon_e^2\epsilon_B^{1/2}
\left({t_{dur}\over10~{\rm s}}\right)^{-3/4}\xi^{-3/2}l_{18}^{3/4}n_1^{1/2}.
\end{equation}
In the case of the reverse shock, 
IC scattering has the possibility of affecting the cooling time
and thereby the efficiency.  If $\epsilon_e<\epsilon_B$, we have
seen that the $y$ parameter is less than unity and the overall efficiency
is just given by
\begin{equation}
\label{effireverse}
\epsilon_{tot,r}=
\epsilon_e\epsilon_{syn}=0.19\epsilon _e^{3/2}\epsilon _B^{1/8}
\left({t_{dur}\over10~{\rm s}}\right)^{-3/16}\xi^{-3/8}
l_{18}^{3/16}n_1^{1/8},
\qquad \epsilon_e<\epsilon_B.
\end{equation}
However, if $\epsilon_e>\epsilon_B$, then $y=(\epsilon_e/\epsilon_B)^{1/2}
>1$ and the efficiency is further reduced to
$$
\epsilon_{tot,r}={\epsilon_e\epsilon_{syn}\over y}
=0.19\epsilon _e\epsilon _B^{5/8}
\left({t_{dur}\over10~{\rm s}}\right)^{-3/16}\xi^{-3/8}
l_{18}^{3/16}n_1^{1/8},
\qquad \epsilon_e>\epsilon_B.
$$

We can see from these expressions that efficient bursts are possible
only if $\epsilon _e$ is high.  The value of $\epsilon _B$ is
important, as far as efficiency is concerned, only if $\epsilon_e
>\epsilon_B$, and that too only for the reverse shock.  We give some
specific estimates of the efficiency in different regions of the
parameter space in the following sections.

\section{Exploration of the parameter space}

We have introduced in the previous sections several parameters. Two
of these are relatively well known, namely the ISM density $n_1 \sim
1~{\rm cm^{-3}}$ and the explosion energy $E$ which determines the
Sedov length scale $l\sim 10^{18}$ cm.  Four other parameters are
essentially unknown.

Two ``external parameters" depend on the properties of the initial
fireball and are likely to vary from one explosion to the next:

1) The Lorentz factor of the shell $\gamma $.

2) The shell thickness $\Delta$, which we prefer to replace by the
parameter $\xi$ defined in equation \ref{xidef}.  As already explained,
$\xi<1$ for a relativistic reverse shock and $\xi=1$ for a spreading
Newtonian shell.  

\noindent
The above two parameters determine the Lorentz factor $\gamma_2=\gamma
\xi^{3/4}$ of the shocked material and the hydrodynamic time (Eq.
\ref{tnr}) on which the kinetic energy of the expanding shell is
converted to thermal energy.  In the case of the forward shock
$\gamma_2$ also determines the thermal energy density and the particle
density of the post-shock gas.  Therefore, all observables from the
forward shock are functions only of $\gamma_2$ rather than of $\gamma$
and $\xi$ individually.  We prefer to express our results in terms of
the observed burst duration $t_{dur}$ rather than $\gamma_2$, using
equation (\ref{g2}) to convert from one to the other.  The reverse shock
is more complicated as the results depend in general on both $\gamma$
and $\xi$.  We choose to express our results as functions of $t_{dur}$
and $\xi$, using equations (\ref{xidef}) and (\ref{tnr}) to transform
from $\gamma$, $\xi$.

In addition to the above two ``external parameters'' there are two
``internal parameters" which depend on the unknown details of the
microphysics in the shocked regions:

3) The energy density in the magnetic field which we describe by
means of the field equipartition parameter, $\epsilon _B$.

4) The fraction of the thermal energy which goes into the electrons,
$\epsilon_e$.

\noindent
Our current understanding of field amplification and particle
acceleration in ultra-relativistic shocks is quite primitive and it is
not possible to estimate either of these parameters from first
principles.  We treat them therefore as free parameters and try to
deduce their values from the constraints imposed by the GRB
observations.  We do, however, assume that the parameters are
relatively constant from one burst to another and between the forward
and reverse shock.  This assumption may well be wrong, but we make it
in the interests of simplicity.

Over all we have a three dimensional parameter space, $t_{dur}
\epsilon_B\epsilon_e$, for the forward shock, and a four dimensional
parameter space, $t_{dur}\xi\epsilon_B\epsilon_e$, for the reverse
shock.  Figures 2 and 3 summarize all the ideas of the previous
sections for the forward shock and reverse shock respectively.

Figure 2 shows the $\epsilon_B\epsilon_e$ plane for the forward shock
for four choices of the observed burst duration $t_{dur}$.  The
forward shock tends to be very energetic so that even the electrons
with the lowest Lorentz factor $\gamma_{e,min}$ often radiate their
synchrotron emission above the BATSE energy range.  Such cases are
not of interest here because they do not contribute to the BATSE
database.  Taking a nominal cutoff energy of 400 KeV to represent the
BATSE threshold (for example more than half of the 54 bursts
analaysed by Band et al. have their break energy between 100KeV and
400KeV), we show by the shaded regions in Fig. 2 the parameter space
which is inaccessible to BATSE.  If BATSE is to detect radiation from
the forward shock, the parameters must correspond to the unshaded
regions.  In general we see that the forward shock is more likely to
be visible to BATSE in a long burst than in a short burst.

The panels in Fig. 2 show two sets of contours: the solid lines
indicate contours of constant duty cycle, calculated with equation
(\ref{dutycyclesyn}), while the dashed lines indicate contours of
constant efficiency, calculated with equation (\ref{effiforward}).
Recall that the forward shock cools almost exclusively by synchrotron
emission and has very little IC scattering.  For a given $t_{dur}$,
the duty cycle depends only on $\epsilon_B$ and is independent of
$\epsilon_e$.  If we use the typical observed duty cycle of a few per
cent as a constraint, we see that the forward shock needs
$\epsilon_B\sim10^{-2}$.  The efficiency of the forward shock is
primarily a function of $\epsilon_e$, with some weak dependence on
$\epsilon_B$ from the fact that a small fraction of the emitted
radiation may be below the BATSE range (Eq.  \ref{eficeincydis}).
This is usually a small factor since the electrons in the forward
shock are usually very energetic.  If we wish to have reasonable
efficiencies, then the forward shock requires $\epsilon_e$ close to
unity (certainly larger than about 0.1).

For the choice $\epsilon_B\sim10^{-2}$, $\epsilon_e\sim1$, Fig. 2
shows that the forward shock radiation is visible to BATSE only for
somewhat long bursts with $t_{dur}$ greater than about 10 s.  If we
increase $\epsilon_B$ also to be of order unity, then only very long
bursts of 100 s duration or longer would be visible to BATSE.


Figure 3 shows the results corresponding to the reverse shock.  In
this case, we need to specify both external parameters, $t_{dur}$ and
$\xi$, to describe the shock, and therefore we display six panels
covering three values of $t_{dur}$ and two values of $\xi$.  Note that
$\xi=1$ corresponds to a Newtonian reverse shock and $\xi<1$ corresponds
to a relativistic shock. 

The reverse shock is always less energetic than the forward shock and
its synchrotron photons are almost always soft enough to be within the
BATSE range.  The gray ``undetectable" regions are therefore limited
to an extreme corner of the parameter space.  Additionally, the soft
nature of the synchrotron radiation means that the Klein-Nishina
effect is not important for IC scattering.  Whether or not IC is
important is therefore determined solely by the relative magnitude of
$\epsilon_B$ and $\epsilon_e$.  IC is important if $\epsilon
_B<\epsilon _e$ and it is not in the opposite case.  The boundary
between the two regimes in indicated by the thick lines in Fig. 3,
with IC being important below the line.

Above the thick lines, the dependences of duty cycle and efficiency
on $\epsilon_B$ and $\epsilon_e$ are similar to those shown for the
forward shock (Fig. 2).  The only difference is that the efficiency
of the reverse shock is generally lower than that of the forward
shock because the bulk of the synchrotron emission occurs below the
BATSE range.  In the regions below the thick lines in Fig. 3, where
IC is important, the duty cycle is modified relative to the pure
synchrotron case.  Because of this, the solid lines are tilted and
take on a dependence on $\epsilon_e$.  The efficiency too is
influenced by IC scattering since it scatters a considerable part of
the energy above the BATSE range.


As $\gamma$ decreases the reverse shock becomes cooler and it emits
less and less photons into the BATSE range.  The reverse shock is,
therefore, efficient only for short durations ($<0.1\sec$).  As with
the forward shock the value of $\epsilon_B$ must be relatively high
in order to get bursts with short duty-cycles. The demand here is
however a little weaker in the IC zone (below the thick lines).
Similarly, high values of $\epsilon_e$ are needed in order to get
high efficiency.  Indeed, for long duration bursts, even with
$\epsilon_e\sim1$ the efficiency is only about 10\%.

Although Figs. 2 and 3 include almost all the information we need, it
is instructive to look at simpler plots corresponding to more
restricted conditions. Specifically, we set $\epsilon _e=0.5$,
corresponding to half the thermal energy of the shocks going into the
electrons.  This is the largest plausible value for this parameter.
We feel that it is necessary to select such a high value because
burst efficiency is quite a strong constraint and in many source
models the observed GRB energies (within the cosmological scenario)
are barely possible even with $\epsilon_e \sim1$.

In Fig. 4, we show the results for the forward shock in the
$t_{dur}$$\epsilon_B$ plane, while in Fig. 5 we show four panels of
$t_{dur}\epsilon_B$ for the reverse shock.



In general we see that for a given choice of $t_{dur}$ and $\epsilon_B$,
only one of the two shocks is important.  For long durations (which
corresponds to relatively low values of $\gamma$) the reverse shock
is too soft and is inefficient in the BATSE band, and so the BATSE
signal is dominated by the efficient forward shock.  On the other
hand, for short durations (where $\gamma$ is relatively high) the
reverse shock is more efficient while the forward shock is too hard
and radiates outside the BATSE range.  This systematic difference
between the two shocks leads to a possible scenario to explain the
observed bimodal distribution of burst durations.  The scenario is
described in the next section.

Figures 4 and 5 also confirm once again that we need a fairly large
$\epsilon_B$ ($>10^{-2}$).  This requirement comes from two
directions.  First $\epsilon_B>10^{-2}$ is necessary in order to have
a duty-cycle in the forward shock of a few percent.  Secondly, it is
needed if we do not wish to reduce the efficiency of the reverse
shock too much.  Although we could in principle choose a value of
$\epsilon_B$ anywhere between $10^{-2}$ and 1, we prefer to set
$\epsilon_B\sim10^{-2}$ because for this choice the forward shock is
visible to BATSE down to durations $\sim10$ s, whereas with
$\epsilon_B\sim1$ the forward shock is detectable only for $t_{\rm
dur}>30$ s.

\section{Bimodality of Burst Durations}

As an example of the implications of the analysis presented in this
paper, we describe here a scenario which provides a possible
explanation for the observed bimodality of GRB durations.

We assume that all bursts have the same values of $n_{1}$,
$l_{18}$, $\xi$, $\epsilon_B$ and $\epsilon_e$, and that the only
parameter which distinguishes one burst from another is the Lorentz
factor $\gamma$ of the expansion of the relativistic shell.  This is
obviously much too simple.  Nevertheless, even this simple model
provides a nice separation of observed bursts into two classes: short
bursts where BATSE detects radiation only from the reverse shock, and
long bursts where most of the BATSE radiation is from the forward
shock.  In addition, some of the observed differences between short and
long bursts are also explained.

We make the following choices for the values of the fixed parameters:

1) We take $\epsilon _B=10^{-2}$ in both the forward and reverse shock.
This value is high enough to yield the observed duty-cycles and low
enough to allow the forward shock emission to fall within the BATSE
range for long duration bursts.

2) $\epsilon _e=0.5$ in both the forward and reverse shocks. This is in
order to make the bursts as efficient as possible.

3) $\xi =1$, i.e. all bursts correspond to the Newtonian regime for
the reverse shock. This assumption is merely for simplicity so that
we do not need to worry about a second parameter $\xi$, in addition
to $\gamma$.  We expect $\xi=1$ to be satisfied if all bursts begin
with sufficiently thin shells initially, e.g.  $\Delta <10^7$, and
spread out so as to be quasi-relativistic.

4) $l=10^{18}$ cm, $n_1=1~{\rm cm^{-3}}$, i.e. all the bursts have
the same explosion energy and expand into a standard ISM.

In this model, the distributions of various burst properties are
determined by the distribution of $\gamma$ over the burst population.
This distribution could in principle be a complicated function, but
we make the simple assumption that $\gamma $ is uniformly distributed
in logarithm from low values of $\gamma$ up to some upper limit
$\gamma_{max}\sim10^4$.  We further assume (again for simplicity)
that the bursts originate in a Euclidean space. This is almost
certainly wrong since there is evidence for a considerable
cosmological effect in the BATSE data.  Nevertheless, the assumption
is appropriate for such a crude model.  The Euclidean assumption
allows us to estimate the number of bursts expected to be detected
simply as the detection-efficiency to the power of $3/2$.

Using the above assumptions we can calculate the relative numbers of
bursts expected to be detected by BATSE as a function of burst
duration $t_{dur}$.  Each value of $t_{\rm dur}$ corresponds to a
fixed value of $\gamma$ (see eq. 3 with $\xi=1$ and fixed $l_{18}$).
For this $\gamma$, we calculate the synchrotron emission as a function
of the observed photon energy $h\nu_{obs}$ separately for the forward
and reverse shock.  We then calculate the photon count rate which
BATSE would detect from this burst and take the 3/2 power of this
quantity to obtain the relative volume over which BATSE is sensitive
to such bursts.  We say that the number of bursts expected to be
detected is proportional to the volume.  Figure 6 shows the calculated
distribution of detected bursts as a function of burst duration.  The
sudden break at $t_{\rm dur}\sim10$ s arises because for shorter
bursts the radiation from the forward shock is too hard to fall within
the BATSE range.


Although the above calculation gives the main idea, some further
details related to BATSE's detection criteria need to be taken into
consideration. BATSE detects a burst if the photon count rate is
higher than a limit set by the background.  This limit varies
depending on the particular trigger used.  BATSE's triggers are set
by the count rates in three time resolutions, $64$ ms, $256$ ms and
$1024$ ms.  The count rate threshold is lower in the $1024$ ms
channel by a factor of 2 than in the 256 ms channel, and by a factor
of 4 compared to the 64 ms channel.  Thus, long bursts are detected
most sensitively by the 1024 ms channel.  On the other hand, short
bursts are preferentially detected in the shorter channels because
their signal becomes smeared out in the 1024 ms channel.  It is
straightforward to include this selection effect in the model (see
Mao, Narayan \& Piran 1994).  Figure 7 shows the expected
distribution of durations of detected bursts with the correction
included.


We see that Fig. 7 qualitatively resembles the observed distribution
of burst durations except for a few differences which are expected and
explained below.  The model predicts the number of bursts to grow with
decreasing duration even below $10$ ms, whereas the observed
distribution seems to roll over at short durations.  This discrepancy
is easily fixed if we take the distribution of $\gamma$ not to be
uniform in $\log\gamma$ but to have fewer bursts with high $\gamma$
(as is plausible).  The model also has an abrupt step at $t_{dur}\sim
10$ s caused by the fact that the forward shock is not detected for
shorter duration.  The step would be rounded off if we do not take the
various parameters other than $\gamma$ to be absolute constants but
allowed them some variation.  The shape of the peak would change also
if we allow electrons to cool below $\gamma_{e,min}$ (see the comment
below equation 23), but we ignore this effect in the present
simplified discussion.

One interesting result of the above exercise is that the bursts in the
second peak, the long bursts, have efficiencies close to unity and
therefore would correspond to a BATSE-observed luminosity of order
$E\sim10^{51}$ ergs.  However, the short bursts generally have
significantly lower efficiencies of order a few percent.  Therefore,
this model predicts that the short bursts should be less luminous, and
therefore that BATSE should detect them out to a shorter distance than
the long bursts.  Mao et al. (1994) found exactly this effect in the
data when they carried out a detailed analysis of the BATSE sample of
bursts.  Further, Mao et al. showed that the long and short bursts
have similar peak luminosities even though they differ by a factor of
30 in their durations and fluences.  Our model does indeed produce
similar peak luminosities within the BATSE band for the two classes of
bursts.  For example a burst of duration $10$ s has the same
luminosity as a short burst of $0.2$ s while a burst of duration $2$ s
which is too short for the forward shock to be detected is less
luminous by a factor of about $15$.

\section{Discussion}

Our basic picture follows the scenario proposed by M\'esz\'aros \&
Rees (1992), namely that a GRB is produced when a relativistic outflow
of particles from a central explosion is slowed down by interaction
with an external ISM.  The interaction takes place in the form of two
shocks---a forward shock which propagates into the ISM and a reverse
shock which propagates into the relativistic shell.  We have used the
results of Sari \& Piran (1995) to express the density, velocity, and
thermal energy of the gas in the shocked regions in terms of physical
parameters such as the Lorentz factor $\gamma$ of the relativistic
flow and the density of the external medium.  Some properties of the
shocked medium are, however, impossible to estimate from first
principles.  One such is the energy density of the magnetic field,
which we write as a fraction $\epsilon_B$ of the total energy.
Another is the fraction of the thermal energy which goes into
electrons, which we write as $\epsilon_e$.  We consider $\epsilon_B$
and $\epsilon_e$ to be free parameters (though constrained to be less
than unity) which we adjust by comparing the predictions of the model
with observations.

We calculate the radiation emitted by the shocked gas via synchrotron
emission and also consider modifications introduced by inverse
Compton (IC) scattering.  In comparing the predictions of the model
to observations, we use two important constraints.  First, we know
that most GRBs have complex time structure with a duty cycle (defined
to be the duration of the sharpest feature divided by the overall
duration of the burst) of a few percent.  We therefore require the
cooling time of the shock-heated electrons in our model to be short
enough to satisfy this constraint.  Second, we impose the requirement
that a reasonably large fraction of the shock-generated thermal
energy should ultimately be visible as radiation within the BATSE
window, 25 KeV to 1 MeV.  Most proposed models of GRBs have a limited
overall energy budget $\sim10^{53}-10^{54}$ ergs and convert only a
percent or so of this energy into kinetic energy of the expanding
shell.  If there were any further inefficiency in converting the
kinetic energy into BATSE-visible radiation then it would be very
hard to match the observed fluences of GRBs, which correspond to a
$\gamma$-ray energy output $\sim10^{51}$ ergs per burst.

One of the primary results of this paper is that we have calculated
the cooling time of the shock-heated electrons via synchrotron
emission and IC scattering.  We reach several interesting
conclusions.

\noindent
1. For photons detected by BATSE, say of energy 100 KeV, there is a
very well-defined relation between the synchrotron cooling time
$\tau_{\rm syn}$ of the electrons and the observed duration of a GRB
$t_{\rm dur}$.  The relation is given in equation (\ref{tausyn2}) and
is the same for both the forward and reverse shocks, and is
independent of whether the reverse shock is Newtonian or
relativistic.  Equation (\ref{dutycyclesyn}) gives a formula for the
duty cycle of a burst and predicts that the duty cycle should be
nearly independent of burst duration.  This prediction appears to be
supported by observations but needs to be checked against the data in
more detail.

\noindent
2. If we require the duty cycle to be no more than a few percent, as
suggested by the observations, then we find that the magnetic field
parameter $\epsilon_B$ must be greater than about $10^{-2}$ (assuming
$\epsilon_e\sim1$, see below).  Such strong fields are not expected
merely from flux-freezing, especially in the forward shock where the
field in the external ISM is likely to be very low.  We conclude,
therefore, that the ultra-relativistic shocks in GRBs must have some
mechanism to build up the magnetic field to near-equipartition
strength in the post-shock gas.

\noindent
3. We find that the duty cycle should decrease with increasing photon
energy as $(h\nu_{\rm obs})^{-1/2}$.  Such a variation in the widths
of features in GRB light curves has been reported by Fenimore et al.
(1995).  These authors find a variation of the form $(h\nu_{\rm
obs})^{-0.4}$, which is reassuringly close to our predicted scaling.

\noindent
4. IC scattering does not modify the results for the forward shock
because the scattering cross-section is strongly suppressed by the
Klein-Nishina effect.  For the reverse shock, IC can be important in
some cases.  While the IC radiation never falls within the BATSE
window, the electrons which produce the BATSE-visible synchrotron
radiation can be significantly cooled by IC if $\epsilon_B <
\epsilon_e$.

In addition to the cooling time scale, we also calculate the
efficiency of a burst.  The efficiency depends on the parameter
$\epsilon_e$, which determines how much energy is available in the
electrons, but also on exactly how much of the energy is actually
radiated within the BATSE range.  Our analysis leads to the following
conclusions.

\noindent
1. We find that the radiation from the forward shock is visible to
BATSE only for bursts with long durations, $t_{\rm dur}>10$ s.  In
shorter bursts, the radiation from the forward shock is too hard and
falls in the MeV range.  This implies that there is a populations of
MeV bursts to which BATSE is not sensitive and which may be worth
searching for in future missions (see Piran \& Narayan 1995).

\noindent
2. The radiation from the reverse shock falls within the BATSE window
for all bursts.  However, the amount of radiation received is maximum
for very short duration bursts and decreases with increasing burst
duration.

\noindent
3. Combining the above two results, and imposing the requirement of
reasonable burst efficiency, we conclude that $\epsilon_e$ must be
large.  This means that ultra-relativistic shocks must be able to
accelerate electrons almost as effectively as they accelerate ions.
For quantitative estimates, we choose $\epsilon_e=0.5$, which
corresponds to the shock thermal energy going into ions and electrons
in equal amounts.  Even with this choice, we find that the reverse
shock is fairly inefficient, with efficiencies as low as 1\% for
$\epsilon_B=10^{-2}$ and $t_{\rm dur}>1$ s.  If we give up burst
efficiency as a criterion, then some of the constraints stated
earlier become looser.  For instance, it may be possible to accept
values of $\epsilon_B$ as low as $10^{-5}$ (as in M\'esz\'aros, Laguna \&
Rees 1993).

\noindent
4. For the particular choice of parameters we favor,
$\epsilon_B=10^{-2}$, $\epsilon_e=0.5$, the model provides a natural
explanation for the bimodal distribution of burst durations.  Long
bursts are produced by the forward shock from fireball shells with
somewhat low values of $\gamma\sim100$.  These bursts are efficient.
Short bursts, on the other hand, are produced by the reverse shock
from higher $\gamma$ events.  These are generally much less efficient.
The model even explains the curious feature noted by Mao et al. (1994)
that the {\it luminosities} of short and long bursts are similar even
though their fluences differ by a large factor.  This results
naturally in our model from the different efficiencies of the two
shocks.

We conclude with two caveats.  First, our model in its present form
does not have an explanation for the break seen by BATSE in many GRB
spectra at around a few hundred KeV.  In the present paper we merely
required that the model should be able to produce photons visible to
BATSE and we calculated the properties of these photons and the
electrons which produce them.  An additional requirement we could have
imposed, but did not, is a low energy cut-off so that the model does
not produce an excess of X-ray photons.  The forward shock in our
model does naturally have a low energy cutoff, and is possibly
consistent with measured constraints in the X-ray band, but the
reverse shock in our model produces too much low energy radiation.  It
is possible that IC cooling suppresses the lower energy radiation.  To
investigate this, the IC interactions will need to be calculated in
greater detail than we have done in this paper.

The second point is that we have assumed that the overall duration of
a burst $t_{\rm dur}$ is given by the hydrodynamical time $t_{\rm
hyd}$, and that this time is greater than the cooling time $t_{\rm
cool}$ of the electrons.  We associate the widths of individual
features in burst profiles with $t_{\rm cool}$.  We find this choice
natural.  However, it is possible to consider the oposite case in
which $t_{\rm hyd}<t_{\rm cool}$.  In such a scenario, individual
spikes in the burst profile would correspond to $t_{\rm hyd}$ and the
overall duration of the burst and its complicated time structure would
need to be generated by some other mechanism, perhaps time variable
ejections in the original source (e.g. Narayan et al. 1992).  One
problem with such a model is that the efficiency will be low.  The
expanding shell slows down on the hydrodynamic time and so any
radiation produced after a time $t_{\rm hyd}$ is no longer
Lorentz-boosted and will therefore arrive over a very much longer time
than $t_{\rm hyd}$ in the observer frame.  Therefore, only a fraction
$t_{\rm hyd}/t_{\rm cool}$ of the emitted radiation will be counted as
being part of the GRB, leading to loss of efficiency.  On the other
hand, the model might be able to explain the low-energy spectal
indices of GRBs and the paucity of X-ray photons (Katz 1994).

\bigskip
This work was supported in part by NASA grant NAG 5-1904 and by an
Israel-NSF grant.

\vfill\eject

\noindent
{\bf Figure Captions}

\bigskip\noindent
Figure 1: Schematic representation of the four zones that are present
when a relativistic fireball interacts with the ISM.  The solid line
indicates the density as a function of radius.  The undisturbed ISM is
zone 1 at large radius, and the unshocked shell material is zone 4 at
small radii.  The shocked zones, 2 and 3, are the result of the
forward and reverse shocks and are separated by a contact
discontinuity.  (Based on Sari \& Piran 1995)

\bigskip\noindent
Figure 2: Properties of the forward shock for four burst durations:
top left: $t_{\rm dur}=0.1$ s; top right: 1 s; bottom left: 10 s;
bottom right: 100s.  Solid lines show contours of the duty-cycle, and
dashed lines show contours of burst efficiency, as fuctions of the two
parameters, $\epsilon_B$ and $\epsilon_e$.  In the shaded regions,
even the lowest energy electrons in the post-shock medium produce most
of their synchrotron radiation above $400$ KeV in the observer frame.
Such bursts are not visible to BATSE.

\bigskip\noindent
Figure 3: Similar to Fig. 2, but for the reverse shock.  The six
panels are as follows: top left: $t_{dur}=0.001$ s, $\xi=0.1$
($\gamma=5 \cdot 10^4$); top right: $t_{dur}=0.001$ s, $\xi=1$
($\gamma=8000$); middle left: $t_{dur}=0.1$ s, $\xi=1$
($\gamma=8000$); middle right: $t_{dur}=0.1$ s, $\xi=1$
($\gamma=1600$); bottom left: $t_{dur}=10$ s, $\xi=0.1$
($\gamma=1600$); bottom right: $t_{dur}=10$ s, $\xi=1$
($\gamma=280$); All unshaded regions are visible to BATSE.  The thick
lines correspond to $\epsilon_B=\epsilon_e$.  IC scattering enhances
the cooling below this line, leading to loss of efficiency of bursts.

\bigskip\noindent
Figure 4: Contours of duty-cycle (solid lines) and efficiency (dashed
lines) for the forward shock as a function of $\epsilon_B$ and
$t_{dur}$.  The results correspond to the choice $\epsilon_e=0.5$.
BATSE can observe bursts only in the unshaded region of the diagram.

\bigskip\noindent
Figure 5: Similar to Fig. 4, but for the reverse shock, for
$\epsilon_e=0.5$.  The four panels are as follows: top left: $\xi=1$;
top right: $\xi=0.1$; bottom left: $\gamma=10^3$; bottom right:
$\gamma=10^4$.  IC cooling is present in all regions and has been
included in calculating the durations and efficiencies.  The dark
region is not physical since there are no burts of duration less than
$0.3$s with $\gamma=10^3$.

\bigskip\noindent
Figure 6: The number of bursts expected to be detected by BATSE as a
function of burst duration if the distribution of sources is
homogeneous and the detection criterion is set solely by luminosity.
The calculations correspond to $\epsilon_e=0.5$ and $\epsilon_B=0.01$.
For the reverse shock we have taken $\xi=1$ (Newtonian shock).  The
peak on the right corresponds to bursts for which the forward shock is
visible to BATSE, while the peak on the left is due to bursts for
which only the reverse shock is visible.

\bigskip\noindent
Figure 7: Similar to Fig. 6, but including the effect of BATSE's three
time channels, 64 ms, 256 ms, 1024 ms, in calculating the
detectability of bursts.  Note the bimodal distribution of durations.
In this model, short bursts are due to radiation from the reverse
shock in fireballs of high $\gamma$, while long bursts are due to the
emission from the forward shock in low $\gamma$ fireballs.

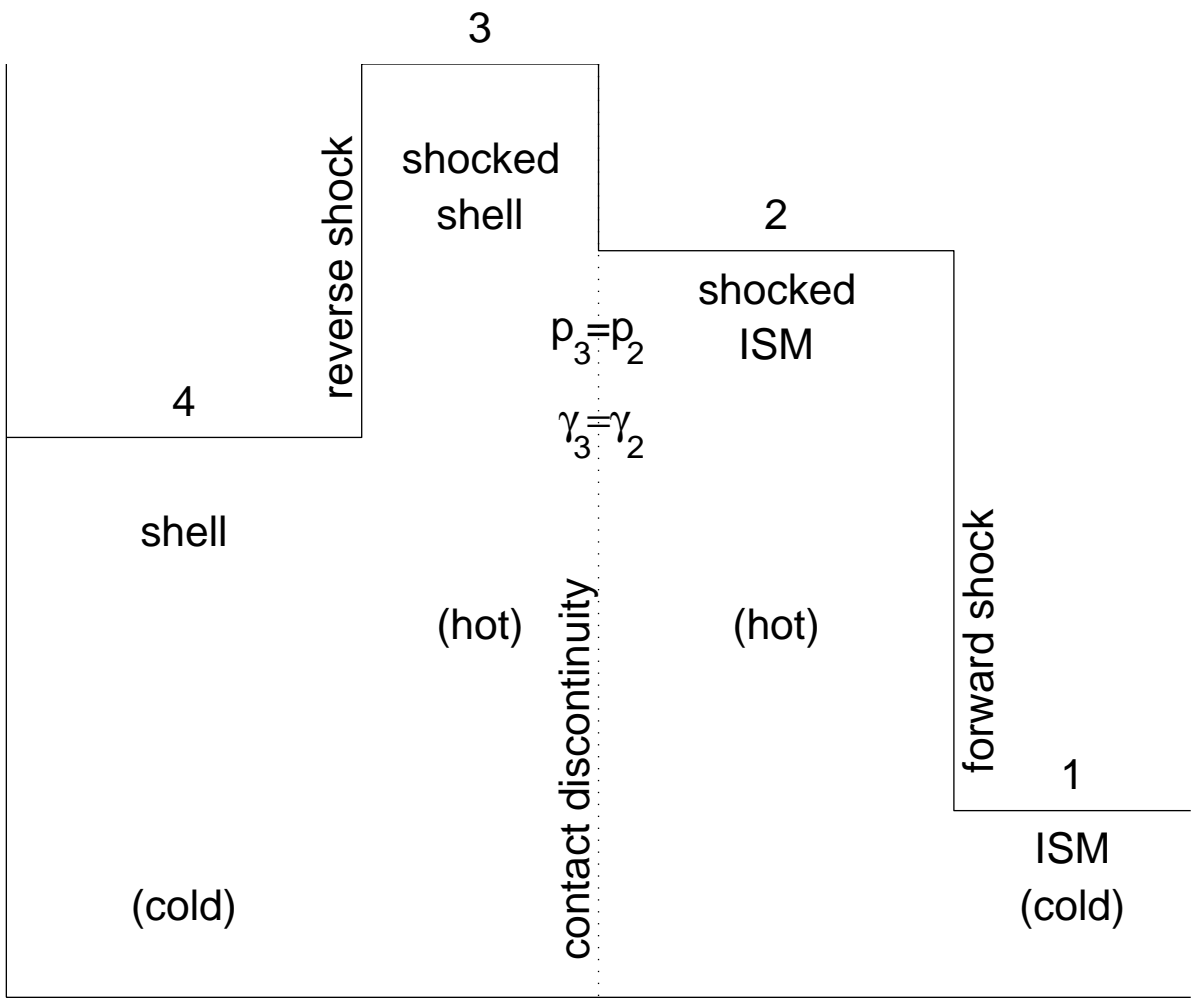
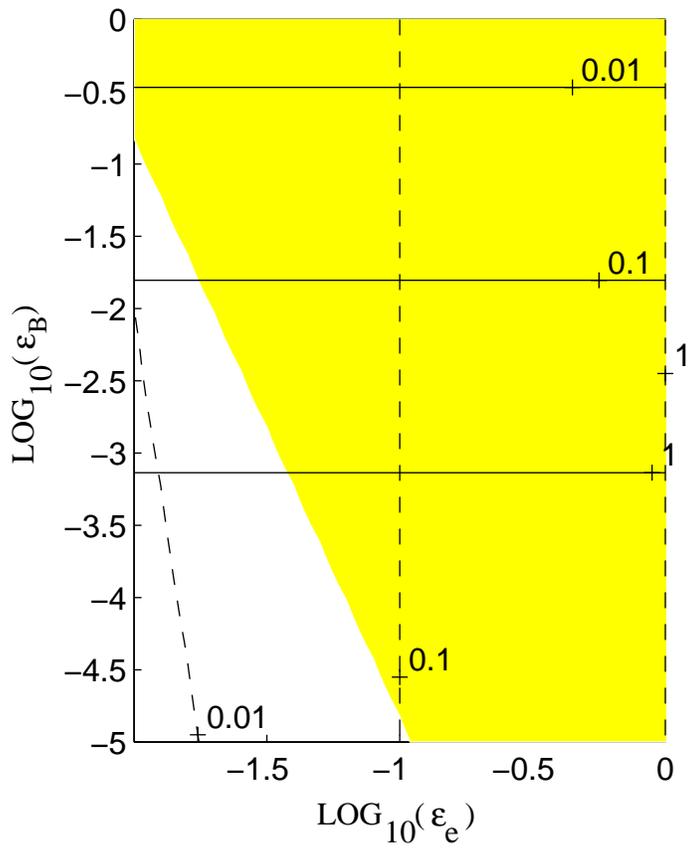
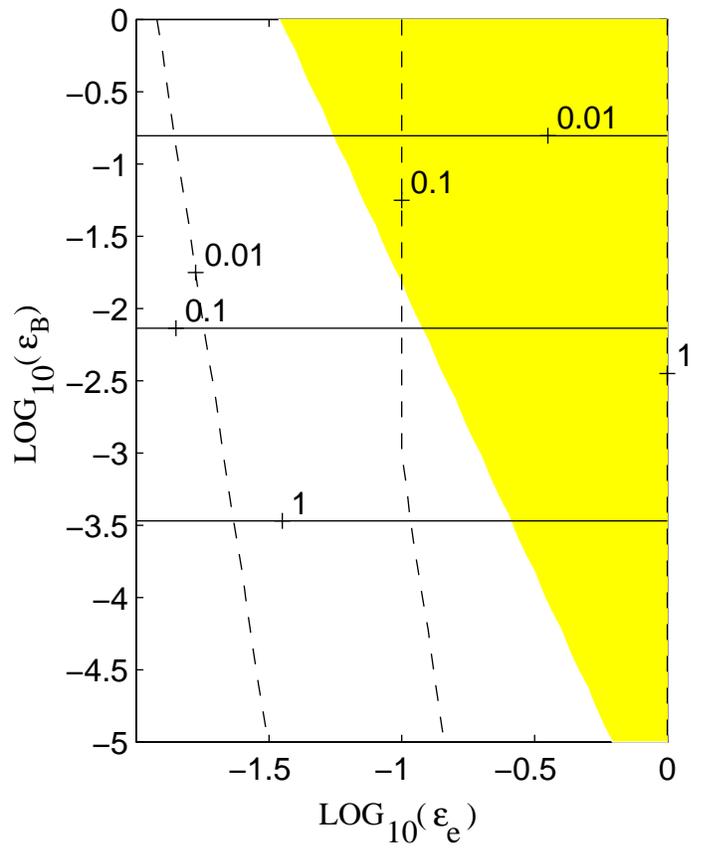
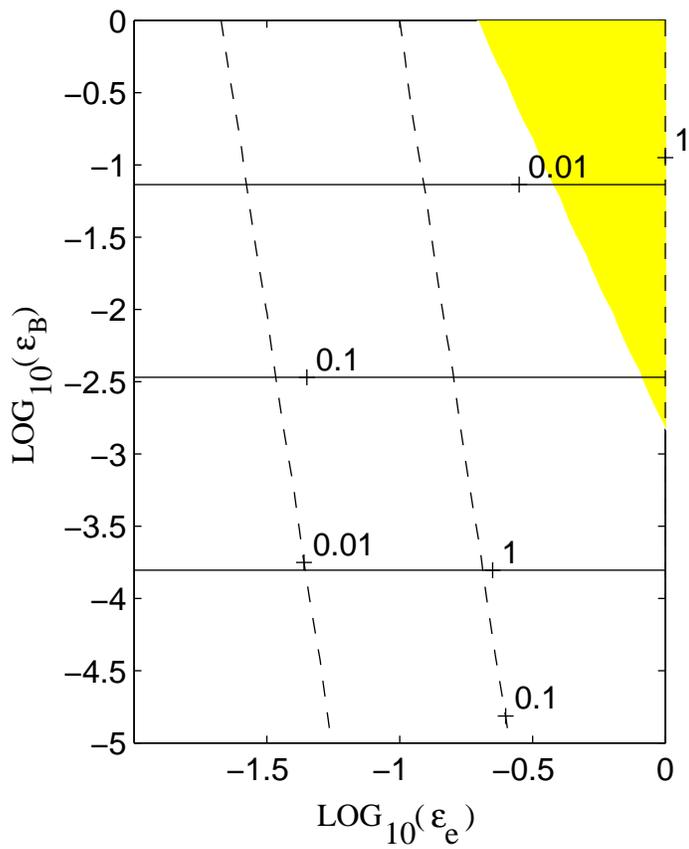
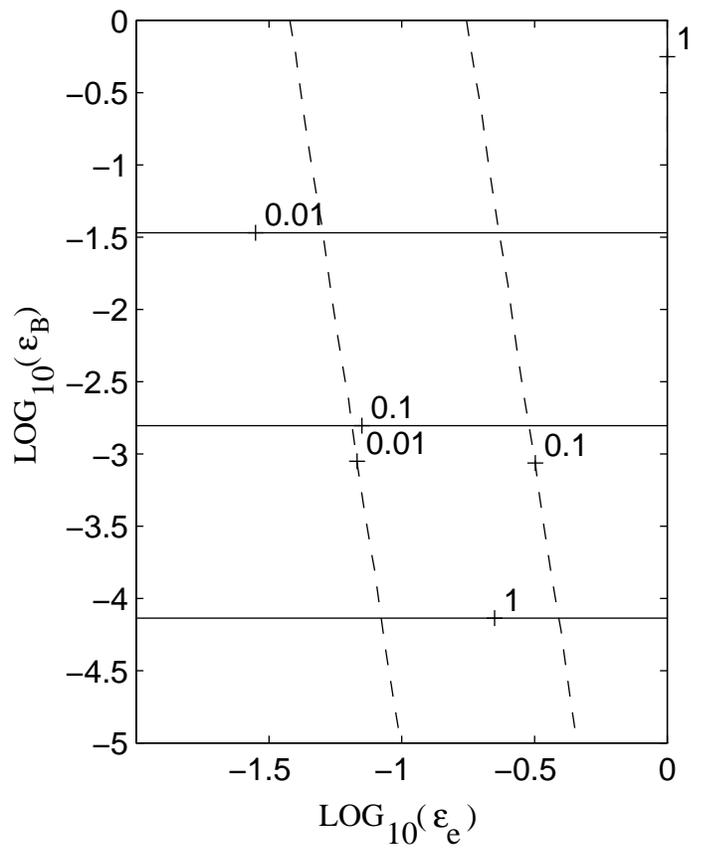
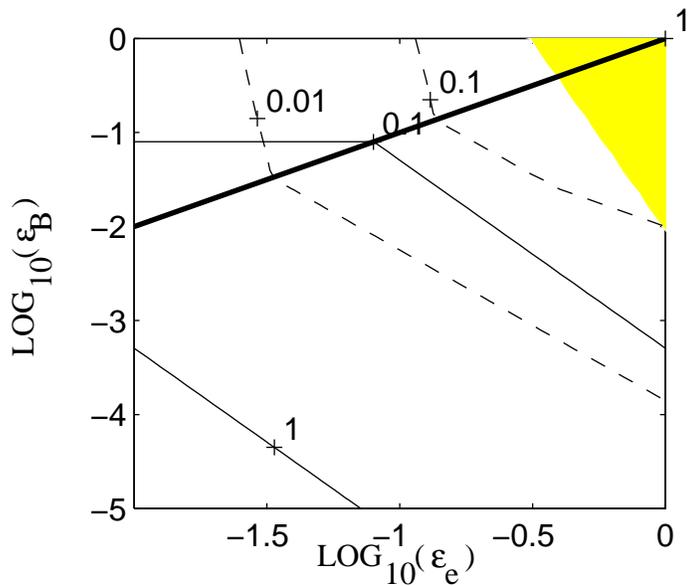
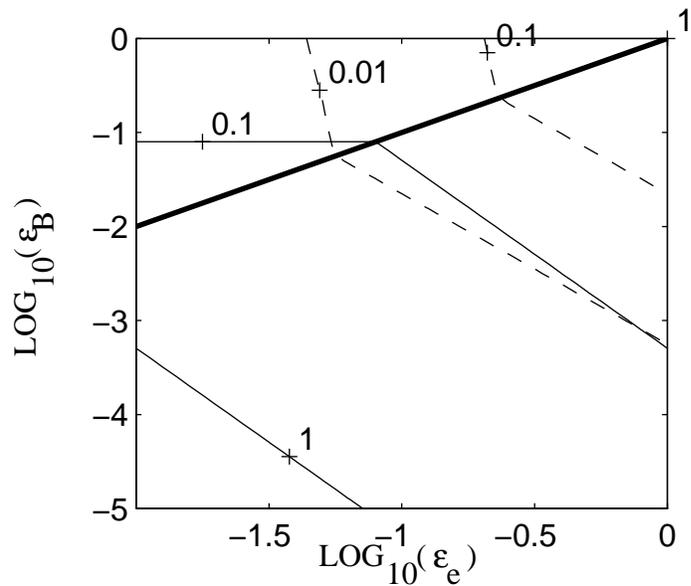
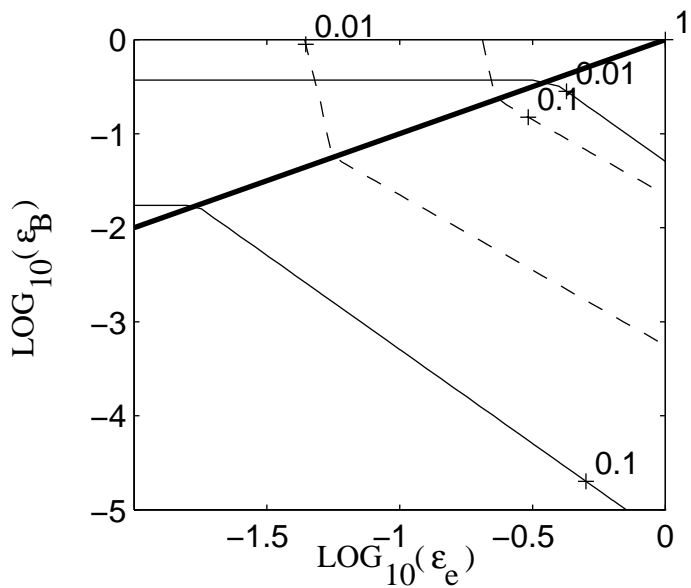
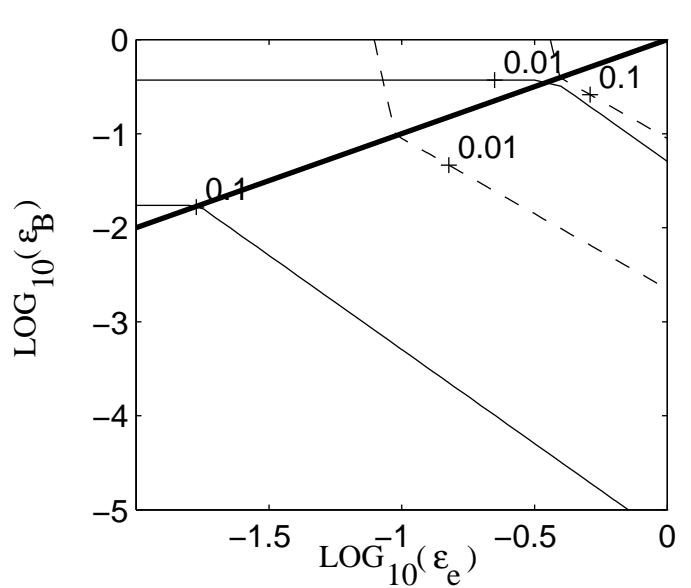
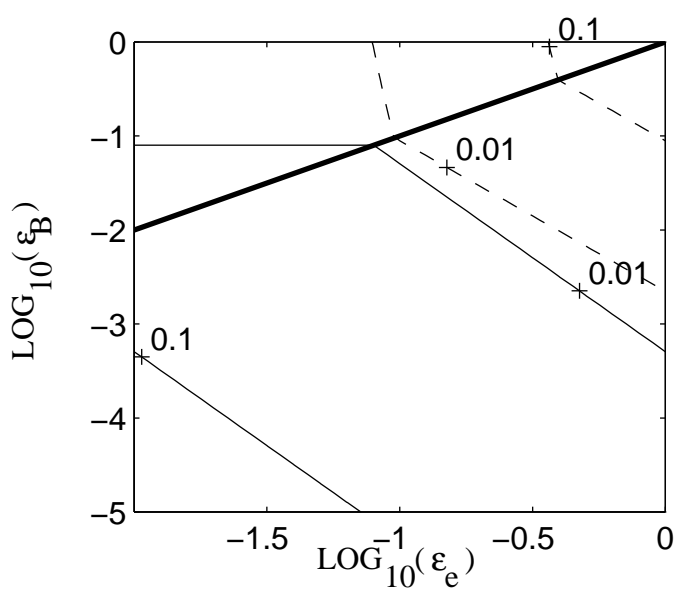
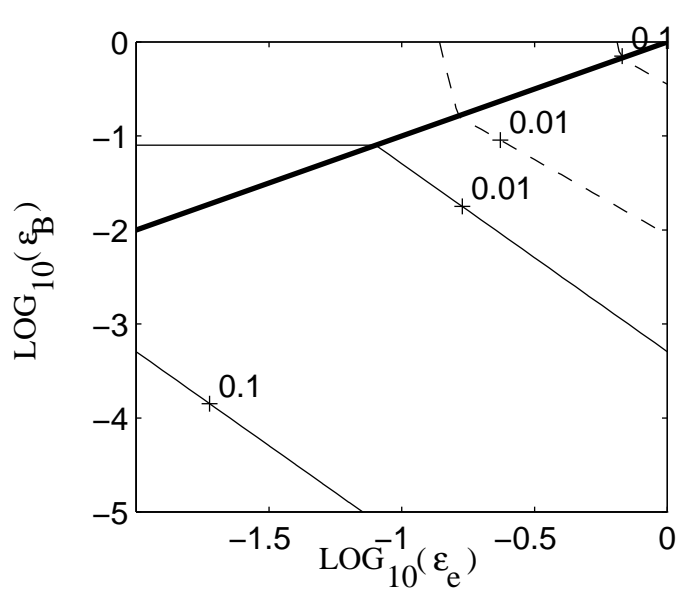
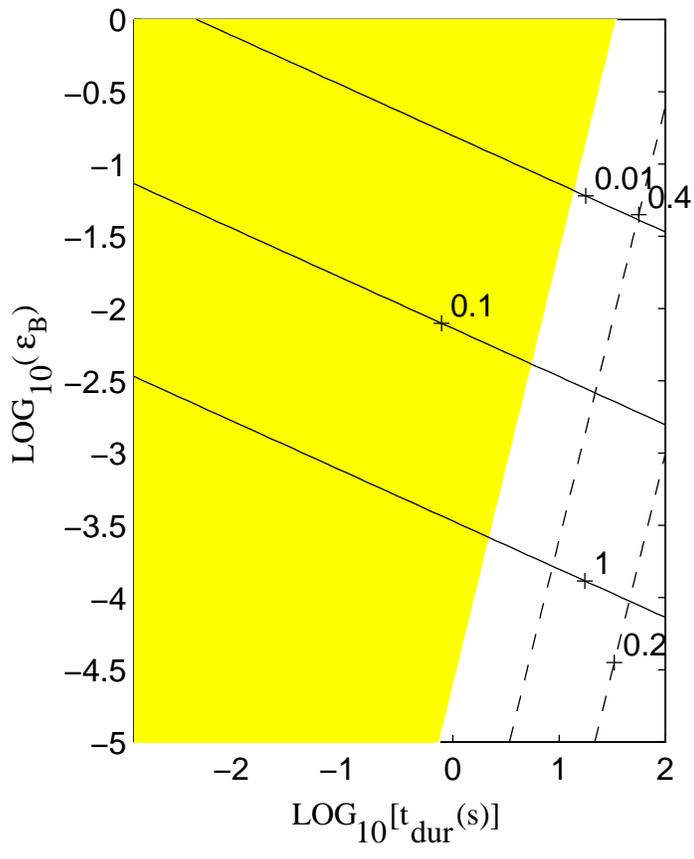
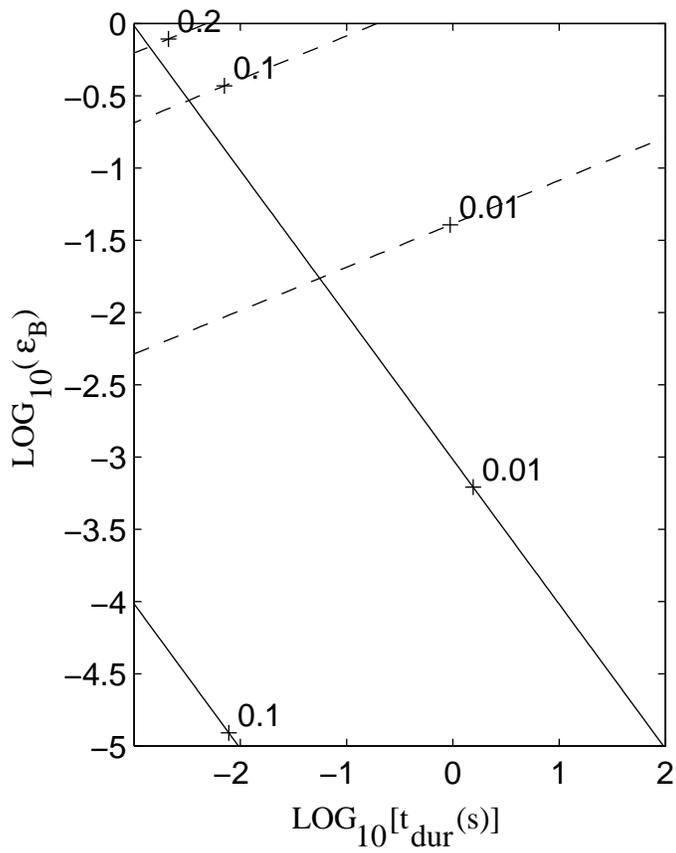
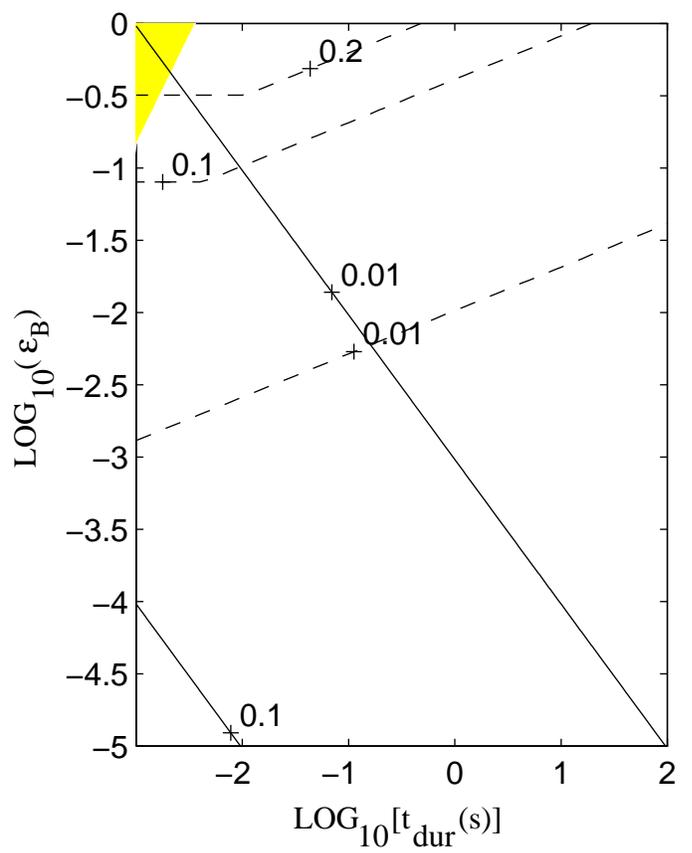
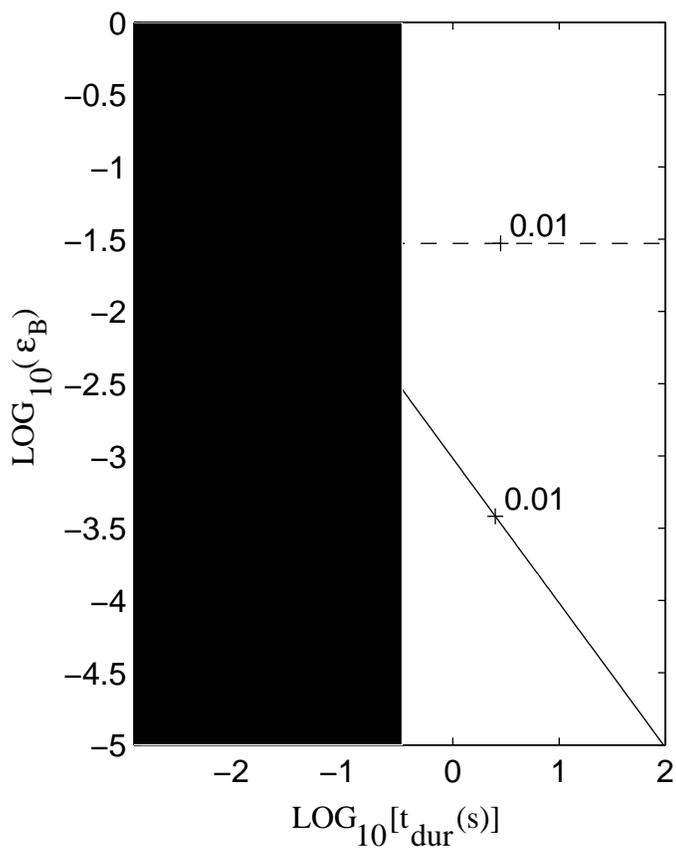
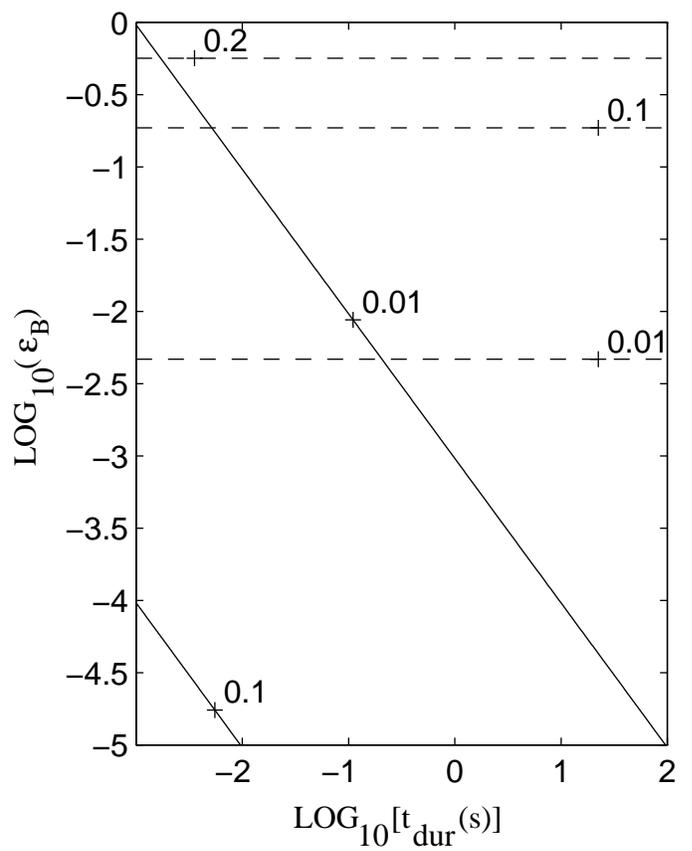
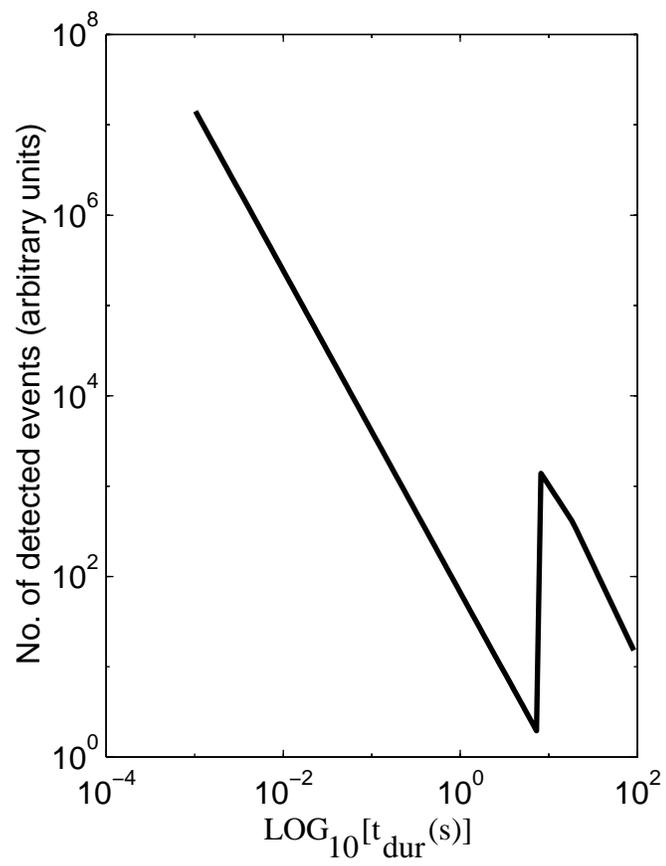
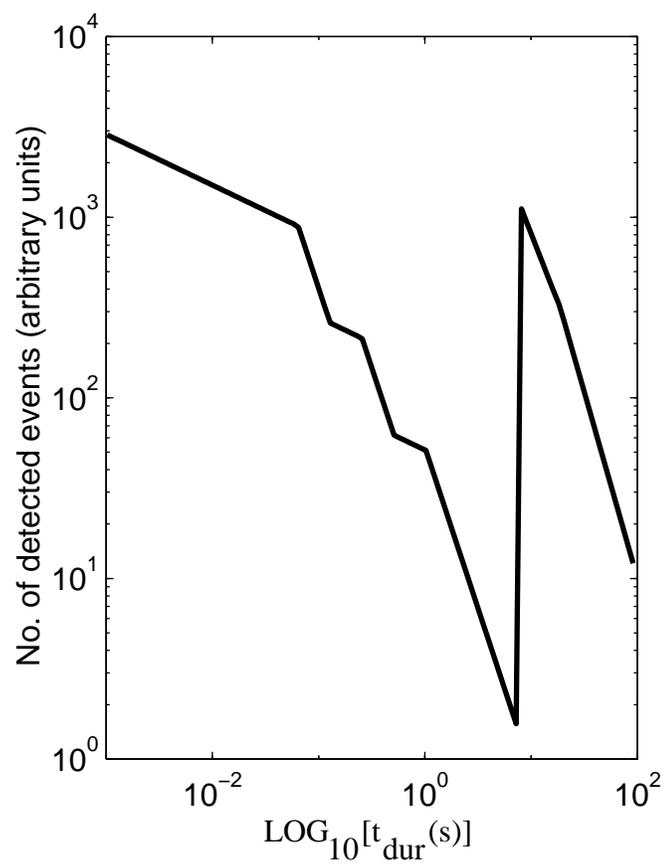
\end{document}